\documentclass[aos]{imsart}
\setattribute{journal}{name}{} %
\setattribute{infoline}{text}{} %

\RequirePackage{amsthm,amsmath,amsfonts,amssymb}
\RequirePackage[numbers]{natbib}
\RequirePackage{graphicx}

\usepackage{enumerate,mathrsfs,comment}
\usepackage{bbm}
\usepackage{enumitem}

\usepackage{mathtools}

\RequirePackage{amsmath}
\RequirePackage{amssymb}
\RequirePackage{amsthm}
\RequirePackage{bm}
\usepackage{bm}
\usepackage{multirow}
\usepackage{graphicx}
\usepackage{subfigure}
\usepackage{makecell}
\usepackage{booktabs}
\usepackage{array}
\usepackage{algorithm}
\usepackage{algorithmic}
\usepackage{dsfont}

\newcommand{\bbE}{\mathbb{E}}

\newcommand{\va}{\boldsymbol{a}}
\newcommand{\vA}{\boldsymbol{A}}

\newcommand{\vB}{\boldsymbol{B}}

\newcommand{\vx}{\boldsymbol{x}}

\newcommand{\vy}{\boldsymbol{y}}

\newcommand{\vz}{\boldsymbol{z}}

\newcommand{\bbeta}{\boldsymbol{\beta}}
\newcommand{\bdelta}{\boldsymbol{\delta}}

\newcommand{\bmu}{\boldsymbol{\mu}}
\newcommand{\btheta}{\boldsymbol{\theta}}

\newcommand{\bOmega}{\boldsymbol{\Omega}}

\newcommand{\bLambda}{\boldsymbol{\Lambda}}

\newcommand{\bSigma}{\boldsymbol{\Sigma}}

\newcommand{\bepsilon}{\boldsymbol{\epsilon}}

\usepackage{scalerel,stackengine}
\stackMath
\newcommand\reallywidehat[1]{%
\savestack{\tmpbox}{\stretchto{%
  \scaleto{%
    \scalerel*[\widthof{\ensuremath{#1}}]{\kern.1pt\mathchar"0362\kern.1pt}%
    {\rule{0ex}{\textheight}}%
  }{\textheight}%
}{2.4ex}}%
\stackon[-6.9pt]{#1}{\tmpbox}%
}

\newcommand{\indep}{\rotatebox[origin=c]{90}{$\models$}}

\newcommand{\cov}{\text{Cov}}

\newcommand{\var}{\text{Var}}

\newcommand{\cS}{\mathcal{S}}

\newtheorem{theorem}{Theorem}

\newtheorem{definition}{Definition}
\newtheorem{remark}{Remark}

\newtheorem{lemma}{Lemma}
\newtheorem{assumption}{Assumption}

\ifx\BlackBox\undefined
\newcommand{\BlackBox}{\rule{1.5ex}{1.5ex}}  %
\fi

\ifx\QED\undefined
\def\QED{~\rule[-1pt]{5pt}{5pt}\par\medskip}
\fi

\ifx\proof\undefined
\newenvironment{proof}{\par\noindent{\bf Proof\ }}{\hfill\BlackBox\\[2mm]}

\fi

\begin{document}

\begin{frontmatter}
\title{On the testing of multiple hypothesis in sliced inverse regression
}

\begin{aug}
  \author{Zhigen Zhao\thanksref{t1}\ead[label=e1]{zhaozhg@temple.edu}}
  \and
  \author{Xin Xing\thanksref{t2}\ead[label=e2]{xinxing@vt.edu}}

  \address{Department of Statistics, Operations, and Data Science, Temple University, Philadelphia, PA, 19122, USA\\ 
           \printead{e1}}
  
  \address{Department of Statistics, Virginia Tech University, Temple University, Blacksburg, VA 24061\\ 
           \printead{e2}}

  \thankstext{t1}{ Zhigen Zhao is Associate Professor of Department of Statistics, Operations, and Data Science, Temple University. }
  \thankstext{t2}{ Xin Xing is Assistant Professor of the Department of Statistics, Virginia Tech University.}

  \runauthor{Zhao and Xing}

\end{aug}

\begin{abstract}
We consider the multiple testing of the general regression framework aiming at studying the relationship between a univariate response and a p-dimensional predictor. To test the hypothesis of the effect of each predictor, we construct an Angular Balanced Statistic (ABS) based on the estimator of the sliced inverse regression without assuming a model of the conditional distribution of the response. According to the developed limiting distribution results in this paper, we have shown that ABS is asymptotically symmetric with respect to zero under the null hypothesis. We then propose a Model-free multiple Testing procedure using Angular balanced statistics (MTA)  and show theoretically that the false discovery rate of this method is less than or equal to a designated level asymptotically. Numerical evidence has shown that the MTA method is much more powerful than its alternatives, subject to the control of the false discovery rate.

\end{abstract}

\end{frontmatter}

\section{Introduction}\label{sec:intro}

In the general framework of the regression analysis, the goal is to infer the relation between a univariate response variable $y$ and a $p\times 1$ vector $\vx$. One would like to know $y|\vx$, namely, how the distribution of $y$ depends on the value of $\vx$. Among the literature of sufficient dimension reduction \citep{li1991sliced, dennis2000save, cook2007dimension, lin2018consistency, lin2019sparse}, the fundamental idea is to replace the predictor by its projection to a subspace without loss of information. In other words, we seek for a subspace $\mathcal{S}_{y|\vx}$ of the predictor space such that  

\begin{equation}\label{eqn:ind}
y\indep \vx |\mathbf{P}_{\mathcal{S}_{y|\vx}}\vx.
\end{equation}
Here $\indep$ indicates independence, and $P_{(\cdot)}$ stands for a projection operator. The subspace $\mathcal{S}_{y|\vx}$ is called the central subspace. Let $d$ be the dimension of this central subspace. Let $\mathbf{B}$, a $p\times d$ matrix, be a basis of the central subspace $\mathcal{S}_{y|\vx}$. Then the equation (\ref{eqn:ind}) is equivalent to 
\begin{equation}\label{eqn:ind:2}
y\indep \vx |\mathbf{B}\vx.
\end{equation}

%
%

To further reduce the dimensionality especially when the number of predictors $p$ diverges with respect to $n$, it is commonly assumed that $y$ depends on $\vx$ through a subset of $\vx$, known as the Markov blanket and denoted as $\mathcal{MB}(y,\vx)$ \citep{pearl1988probabilistic, statnikov2013algorithms, candes2018panning}, such that 
\[
y\indep \vx |\mathcal{MB}(y,\vx).
\]
For each predictor, one would like to know whether $x_j \in \mathcal{MB}(y,\vx)$, which could be formulated as a multiple testing problem. %
The null hypothesis stating that $x_j\notin \mathcal{MB}(y,\vx)$ is equivalent to 
\begin{equation}\label{eqn:hypothesis}
\mathcal{H}_j: \mathbf{P}_{span(x_j)}( \mathcal{S}_{y|\vx} ) = \mathcal{O}_p,
\end{equation}
where $\mathcal{O}_p$ is the origin point in the $p$-dimensional space \citep{cook2004testing}. In other words, it is equivalent to saying that the $j$-th row of the matrix  $\mathbf{B}$ consists of all zeros.

There are many attempts targeting estimating the central subspace in the existing literature on sufficient dimension reduction. The most widely used method is the sliced inverse regression (SIR) which was first introduced in \cite{li1991sliced}. Later, there are many attempts to extend SIR, including, but not limited to, sliced average variance estimation \citep{cook1991sliced, dennis2000save}, directional regression \citep{li2007directional}, and constructive estimation \citep{xia2007constructive}. Nevertheless, most of the existing methods and theories in sufficient dimension reduction focus on the estimation of the central subspace $\mathcal{S}_{y|\vx}$. The result of statistical inference is very limited when $p$ diverges, not to mention the procedure of controlling the false discovery rate (FDR) when testing these hypotheses simultaneously.

The challenge arises from two perspectives. First, among the literature on sufficient dimension reduction, the result on the limiting distribution of the estimator of the central subspace is very limited when $p$ diverges. When $p$ is fixed, \citep{hsing1992asymptotic, zhu2006sliced, Saracco1997asymptotic, li2009dimension} have derived the asymptotic  distribution of the sliced inverse regression. To the best of the authors' knowledge, there are no results on the limiting distribution when $p$ diverges unless assuming the signal is strong and the total number of false hypotheses is fixed \citep{wu2011asymptotic}. %

Second, after the test statistic is determined for each hypothesis, how to combine these test statistics to derive a method that controls the false discovery rate is challenging. Many existing procedures, such as \cite{barber2015controlling, barber2019knockoff, xing2021controlling, sarkar2021adjusting},  work on the (generalized) linear regression models. In \cite{candes2018panning}, the authors considered an arbitrary joint distribution of $y$ and $\vx$ and proposed the model-X knockoff to control FDR. %
However, this method requires that the distribution of the design matrix is known, which is not feasible in many practice.

The study conducted by \cite{ji2014rate} explored variable selection for the linear regression model, assuming the condition of weak and rare signals. It is noted in this paper that the {\it selection consistency} is not possible and allowing for false positives is necessary. While several existing penalization-based methods  on sufficient dimension reduction require imposing {\it uniform signal strength condition} to achieve consistent results (\cite{li2007sparse, lin2019sparse, tan:2020, zhang2014confidence}), this work tackles a more challenging scenario. Specifically, we develop the central limit theorem of SIR, utilizing recent theories on Gaussian approximation \citep{chernozhukov2013gaussian, chernozhukov2017central} without relying on the uniform signal strength conditions. This theoretical result is the first of its kind in the literature on sufficient dimension reduction and is a necessity for simultaneous inference when the effects of some relevant predictors are either moderate or weak.

We proceed by constructing a statistic for each hypothesis based on sliced inverse regression. Applying Gaussian approximation theory, we demonstrate that this statistic is asymptotically symmetric about zero when the null hypothesis holds. We refer to this as a {\bf A}ngular {\bf B}alanced {\bf S}tatistic (ABS). We then develop a single-step procedure that rejects a hypothesis when its ABS exceeds a certain threshold. Additionally, we provide an estimator for the false discovery proportion. For a designated FDR level $q$, we adaptively select a threshold such that the estimated false discovery proportion is no greater than $q$. This method is referred to as the Model-free multiple Testing procedure using Angular balanced statistics (MTA). Theoretical analysis confirms that MTA asymptotically controls the FDR at the $q$-level under regularity conditions. Simulation results and data analysis demonstrate that MTA significantly outperforms its competitors in terms of power while controlling the FDR.

The paper is organized as follows. In Section \ref{sec:theor:sir}, we derive the central limit theorem of SIR when both the dimension $p$ and the number of important predictors $s$ diverge with $n$ using the recently developed Gaussian approximation theory. In Section \ref{sec:mmm}, we construct ABS based on the estimator of SIR and propose the MTA method. It is shown that the FDR of the MTA method is less than or equal to a designated level asymptotically. In Sections \ref{sec:simulation} and \ref{sec:realdata}, we provide numerical evidence including extensive simulations and a real data analysis to demonstrate advantages of MTA. We conclude the paper in Section \ref{sec:conclusion}, and include all technical details in the appendix.

\subsection*{Notation}
We adopt the following notations throughout this paper.
For a matrix $\vA$, we call the space generated by its column vectors the column space and denote it by $col(\vA)$. The element at the $i$-th row and $j$-th column of a matrix $\vA$ is denoted as $A_{ij}$ or $a_{ij}$. The $i$-th row and $j$-th column of the matrix are denoted by $\vA_{i\cdot}$ and $\vA_{\cdot j}$, respectively. The minimum and maximum eigenvalues of $\vA$ are denoted as $\lambda_{min}(\vA)$ and $\lambda_{\max}(\vA)$ respectively.
For two positive numbers $a,b$, we use $a\vee b$ and $a\wedge b$ to denote $\max\{a,b\}$ and  $\min\{a,b\}$ respectively.
We use $c$, $C$, $C'$, $C_1$ and $C_2$ to denote generic absolute constants, though the actual value may vary from case to case.

\section{Gauss Approximaton of SIR}\label{sec:theor:sir}
Recall that $y$ is the response and $\vx$ is a p-dimensional vector. In the literature of sufficient dimension reduction, people aim to find the central subspace $\mathcal{S}_{y|\vx}$ defined in (\ref{eqn:ind}). The sliced inverse regression (SIR) introduced in \cite{li1991sliced} is the first and the most popular method among many existing ones. %

 Assume that the covariance matrix of $\vx$ is $\bSigma$. Let $\bOmega=\bSigma^{-1}$ be the precision matrix of $\vx$. Let $s_j=\#\{k: \bOmega_{jk}\neq 0\}$ and $s=\max_js_j$.
When the distribution of $\vx$ is elliptically symmetric, it is shown in \cite{li1991sliced} that 
\begin{align}\label{eqn:beta-eta}
\bSigma \mathcal{S}_{y|\vx}=col(\bLambda),
\end{align}
where $\bLambda= var( \bbE[\vx|y])$ and $col(\bLambda)$ is the column space spanned by $\bLambda$. 

Given $n$ $i.i.d.$ samples $(y_{i},\vx_{i})$, $i=1,\cdots,n$. To estimate $\bLambda$, divide the data into $H$ slices according to the order statistics $y_{(i)}$, $i=1,\ldots, n$. Let $\vx_{(i)}$ be the concomitant associated with $y_{(i)}$. Note that slicing the data naturally forms a partition of the support of the response variable, denoted as $\mathcal{H}$. Let $\mathcal{P}_h$ be the $h$-th slice in the partition $\mathcal{H}$. Here we let $\mathcal{P}_1 = (-\infty, y_{(\lceil n/H \rceil)}]$ and $\mathcal{P}_H = (y_{( \lceil n/H\rceil *(H-1)+1 )}, +\infty)$.
Let $\bar{\vx}$ be the mean of all the $\vx$'s and $\bar{\vx}_{h,\cdot}$ be the sample mean of the vectors $\vx_{(j)}$'s such that its concomitant $y_{(j)}\in\mathcal{P}_h$ and estimate 
$\bLambda\triangleq var(\bbE[\vx|y])$ by
\begin{equation}\label{eqn:lambda}
\widehat{\bLambda}_{\mathcal{H}}=\frac{1}{H}\sum_{h=1}^{H}(\bar{\vx}_{h,\cdot}-\bar{\vx})(\bar{\vx}_{h,\cdot}-\bar{\vx})^{\tau}.
\end{equation}
The $\widehat{\bLambda}_{\mathcal{H}}$ was shown to be a consistent estimator of $\bLambda$ under some technical conditions \citep{duan1991slicing, hsing1992asymptotic, zhu2006sliced,li1991sliced, lin2018consistency}.  

Alternatively, we could view SIR through a sequence of ordinary least squares regressions. 
Let $f_h(y), h=1,2,\cdots,H$ be a sequence of transformations of $Y$. Following the proof of \cite{yin2002dimension, wu2011asymptotic}, one knows that under the linearity condition \citep{li1991sliced},
\[
Ef(y)\phi(y)\in \mathcal{S}_{y|\vx},
\]
where $\phi(y) = \bSigma^{-1}E(\vx|y)$.
Let $\bbeta_h\in R^{p\times 1}$ be defined as 
\begin{equation*}
\bbeta_h = argmin_{\bbeta_h} E( f_h(y) - \vx\bbeta_h)^2.
\end{equation*}
Assuming the coverage condition \citep{ni2005note, cook2006ni, wu2011asymptotic}, then 
\[
Span(\mathbf{B}) = \mathcal{S}_{y|\vx},
\]
where $\mathbf{B} = (\bbeta_1,\cdots, \bbeta_H) \in R^{p \times H}$. 

Note that different choices of $f_h(y)$ lead to different methods \cite{wu2011asymptotic,dong2021brief}. To name a few here, \cite{yin2002dimension} suggested $f_h(y)=y^h$ where $h\le H$. After slicing the data into $H$ slices according to the value of the response variable $y$, \cite{cook2006ni} suggested $f_h(y)=y$ if $y$ is in the $h$-th slice and 0 otherwise. %
If we choose $f_h(y) = \mathbbm{1}( y \in \mathcal{P}_h)$, this will lead to SIR, which is the main focus in this paper \citep{yin2002dimension, wu2011asymptotic}.  %

After obtaining data $(\vx_i, y_i)$ based on a sample of $n$ subjects, let 
$$
 f_h(\vy)=\mathbbm{1}(\vy \in \mathcal{P}_h) = \left( \mathbbm{1}(y_1 \in \mathcal{P}_h), \mathbbm{1}(y_2 \in \mathcal{P}_h), \cdots, \mathbbm{1}(y_n \in \mathcal{P}_h) \right)^T.
$$ 
Let $\hat{\bbeta}_h$ be defined as 
\begin{equation*}
\hat{\bbeta}_h = argmin_{\bbeta_h} ||f_h(\vy) - \vx^T\bbeta_h||^2=(\vx\vx^T)^{-1}\vx f_h(\vy),
\end{equation*}
or a general form 
\begin{equation}\label{eqn:ols}
\hat{\bbeta}_h = argmin_{\bbeta_h} ||f_h(\vy) - \vx^T\bbeta_h||^2=\frac{1}{n}\hat{\bOmega}\vx f_h(\vy),
\end{equation}
where $\hat{\bOmega}$ is a suitable approximation of the inverse of the Gram matrix $\hat{\bSigma}=\vx^T\vx/n$.
Let 
\begin{equation}\label{eqn:Bhat}
\hat{\vB}= (\hat{\bbeta}_1,\cdots,\hat{\bbeta}_H).
\end{equation}
There are many methods to estimate the precision matrix. As an example, we could consider the one given by the Lasso for the node-wide regression on the design matrix $\vx$ \citep{meinshausen2006high}. Next, we will derive the central limit theorem of SIR when $p$ converges to the infinity.

Our derivation is built upon the Gaussian approximation (GAR) theory recently developed in \cite{chernozhukov2013gaussian, chernozhukov2017central}. Let $\mathcal{P}_H=\cup_{h}\mathcal{P}_h$ be a partition of the sample space of $y$ and $p_h=P(y\in \mathcal{P}_h)$. %
Define
\begin{equation}\label{eqn:beta:tilde}
\tilde{\bbeta}_h=\frac{1}{n}\bOmega\vx f_h(\vy).
\end{equation}
Let $\tilde{\vB}=(\tilde{\bbeta}_1, \cdots,\tilde{\bbeta}_H)$. For $i=1,2,\cdots, n,j=1,2,\cdots, p, h=1,2,\cdots, H$, let 
$z_{ijh}$'s be normal random variables such that 
\begin{itemize}[leftmargin=*]
    \item $Ez_{ijh}=p_h\bOmega_{\cdot j}^TE(\vx|y\in \mathcal{P}_h)$;
    \item $V(z_{ijh}) = p_h\bOmega_{\cdot j}^TE(\vx\vx^T|y\in \mathcal{P}_h)\bOmega_{\cdot j} -  p_h^2(\bOmega_{\cdot j}^TE(\vx|y\in \mathcal{P}_h) )^2$;
    \item $Cov(z_{ijh}, z_{ikh}) = p_h \bOmega_{\cdot j}^TE(\vx\vx^T|y\in \mathcal{P}_h)\bOmega_{\cdot k} - p_h^2 \bOmega_{\cdot j}^TE(\vx|y\in \mathcal{P}_h)\bOmega_{\cdot k}^T \\E(\vx|y\in \mathcal{P}_h)$;
    \item $Cov(z_{ijh_1}, z_{ikh_2})= - p_{h_1}p_{h_2}\bOmega_{\cdot j}^TE(\vx|y\in \mathcal{P}_{h_1})\bOmega_{\cdot k}^TE(\vx|y\in \mathcal{P}_{h_2})$.
\end{itemize}
Note that when $\mathcal{H}_j$ is true and the coverage condition holds, then $Ez_{ijh}=0$ for any $i$ and $h$ (see Remark \ref{remark:z} in the appendix).

Let $\vz_{i\cdot\cdot}$ be a $p$ by $H$ matrix consisting of $z_{ijh}$ where $j=1,2,\cdots, p$ and $h=1,2,\cdots, H$.
Let $\mathcal{A}^{re}$ consist of all sets $A$ of the form
\begin{equation}\label{eqn:A:re}
A = \{\omega \in R^{pH}: a_j\le \omega_j\le b_j,  j=1,2,\cdots,pH\}
\end{equation}
for some $-\infty \le a_j\le b_j\le \infty$. We will develop a bound of the quantity
\begin{equation}\label{eqn:rho}
\rho_n(\mathcal{A}^{re}) = \sup_{A\in \mathcal{A}^{re}}\left| \mathbb{P}(\sqrt{n}\hat{\mathbf{B}}\in A) - \mathbb{P}(\frac{1}{\sqrt{n}}\sum_{i=1}^n\vz_{i\cdot\cdot} \in A)\right|.
\end{equation}

The theoretical investigation requires the following assumptions

\begin{assumption}\label{assump:1}
The design matrix $\vx$ has iid sub-Gaussian rows. In other words, $sup_{||a||_2\le 1}\mathbf{E}exp\{|\sum_{j=1}^pa_jx_{ij}|^2/C\}\le 1$ for some large enough positive constant $C$. %
\end{assumption}

\begin{assumption}\label{assump:2}
There exists two positive constants $c$ and $C$ such that the smallest eigenvalue $\lambda_{min}$ of $\bSigma$ is greater than $c$ and the largest eigenvalue $\lambda_{max}$ of $\bSigma$ is less than $C$.
\end{assumption}

Assumption \ref{assump:1} is assumed in \cite{zhang2017simultaneous}, describing the tail property of the predictors. The bounds of the eigenvalues of $\bSigma$ are commonly assumed when considering the estimation of the covariance matrix and the precision matrix. %
In the sliced inverse regression, the data are sliced according to the response variable $y$. This creates a partition on the support of the response variable. We assume the following condition on the partition. 

\begin{definition}[Distinguishable Partition]\label{def:partition}
For a given $H$, we call $\mathcal{DP}_H(b)$, a collection of all partitions $-\infty=a_0<a_1<\cdots<a_{H-1}<a_H = \infty$ of $\mathbb{R}$, distinguishable if for any partition in $\mathcal{P}_H$,
\begin{itemize}
\item[(D1)] there exists two constants $\gamma_1$ and $\gamma_2$ such that 
\[
\frac{\gamma_1}{H} \le p_h \le \frac{\gamma_2}{H}, \textrm{where $p_h=\mathbb{P}(a_{h-1}\le Y < a_h)$};
\]
\item[(D2)] $\lambda_{min}\left( Cov\left ( \vx \mathbbm{1}( y\in \mathcal{P}_h ), \vx \mathbbm{1}( y\in \mathcal{P}_h ) \right) \right) > b,\forall h=1,2,\cdots, H,$
where $b$ is a constant which does not depend on n and p. 
\end{itemize}
\end{definition}
The condition (D1) on the partition requires that the probability data falls in each slice is inversely proportional to the number of slices. This condition is assumed in \cite{lin2018consistency} when establishing the phase transition of SIR. The condition (D2) requires a sufficient variation of the $\vx$, which falls in each slice. Otherwise, if the smallest eigenvalue of that covariance matrix in a given slice converges to zero, the corresponding estimator becomes unstable. In Remark \ref{remark:distinguishable}, we give examples that (D2) holds.

\begin{theorem}\label{thm:asym:normal}
Assume Assumptions (\ref{assump:1}-\ref{assump:2}) and a partition $\mathcal{P}_H\in \mathcal{DP}_H(b)$. %
If $s^3p^3\log(pHn)^7/n \le C_1n^{-c_1}$ for some constants $c_1, C_1>0$, then with probability at least $1-2\exp(-cn^\alpha)$, 
\begin{equation}\label{eqn:asym:normal}
\rho(\mathcal{A}^{re})  \le CD_{n,p},
\end{equation}
where $C$ is a constant and $D_{n,p} = \left( \frac{\log^7(pHn)}{n}\right)^{1/6}$.
\end{theorem}
The proof is put in the appendix. From this theorem, it is seen that the estimator $\hat{\mathbf{B}}$ could be approximated by Gaussian random vectors which preserve the mean and covariance structure.

\begin{remark}\label{remark:signal}
When the dimension $p$ is fixed, it is known that the limiting distribution of the estimator based on SIR is normally distributed (\cite{hsing1992asymptotic, Saracco1997asymptotic, zhu1996asymptotics}. In \cite{wu2011asymptotic}, they have used the penalization method to estimate the central space. They have further established the central limit theorem. However, this theoretical development requires a {\it uniform signal strength condition} that the minimum of the magnitude of nonzero parameters is greater than some bound such that the nulls and alternatives could be well separated (\cite{zhang2014confidence}). On the contrary, Theorem \ref{thm:asym:normal} does not rely on the uniform signal strength condition. 

Additionally, this theorem provides a result on the uniform convergence which is essential for simultaneous inference when the effects of some relevant predictors are either moderate or weak. 
\end{remark}

\begin{remark}\label{remark:distinguishable}Comments on the distinguishable condition.
The condition (D2) guarantees that the ordinary least estimation in each slice is stable. This condition holds in examples such as the inverse regression model suggested by Cook \citep{cook2005sufficient, cook2007fisher}. In this model, it is assumed that the covariate $\vx$ is given as 
\[
\vx_y = \mu + \boldsymbol{\Gamma}\boldsymbol{\nu}_y + \sigma\bepsilon,
\]
where $\mu\in \mathbb{R}^p$, $\boldsymbol{\Gamma}\in \mathbb{R}^{p\times d}$, $\boldsymbol{\Gamma}^T\boldsymbol{\Gamma}=I_d$. Then 
\begin{eqnarray*}
&&Cov( \vx\mathbbm{1}(y\in \mathcal{P}_h, \vx\mathbbm{1}(y\in \mathcal{P}_h) ) \succcurlyeq p_h\sigma^2\cov(\bepsilon,\bepsilon).
\end{eqnarray*}

In \cite{jiang2014variable}, the authors considered the following model 
\[
\vx|y\in \mathcal{S}_h \sim MVN(\mu_h, \bSigma), \forall h.
\]
It is easily seen that 
\begin{eqnarray*}
&&Cov( \vx\mathbbm{1}(y\in \mathcal{P}_h, \vx\mathbbm{1}(y\in \mathcal{P}_h) ) \succcurlyeq p_h\bSigma.
\end{eqnarray*}
\end{remark}

\section{Model Free Multiple Testing using Angular Balanced Statistics (MTA)}\label{sec:mmm}
In  this section, we consider testing the hypothesis defined in (\ref{eqn:hypothesis}) simultaneously such that the false discovery rate is controlled at a designated level $q$. We start with the construction of the test statistic.

\begin{figure*}[ht]
\begin{center}
\includegraphics[width = \textwidth]{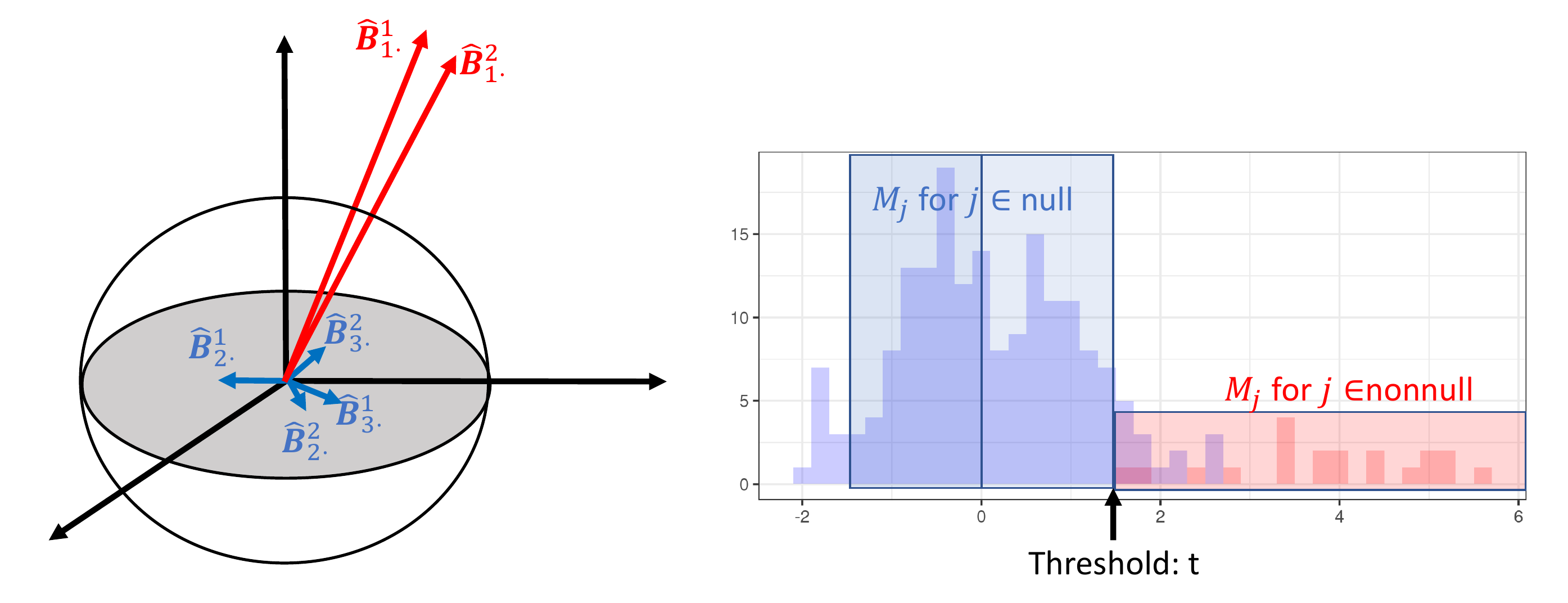}
\caption{In the left panel, we show an illustration of the coefficient vectors for three variables $X_1, X_2$ and $X_3$ where $\mathcal{H}_1$ is false and $\mathcal{H}_2, \mathcal{H}_3$ are true. %
The right panel shows the histogram of mirror statistics for all variables when the data is generated according to Setting 1 with $\rho=0.4$ in Section \ref{sec:simulation}.
} \label{fig:demo}
\end{center}
\end{figure*}

First, we randomly split the sample as two parts, denoted as $\mathcal{D}_1$ and $\mathcal{D}_2$ \citep{dai2020false}. Let $\hat{\mathbf{B}}^1$ and $\hat{\mathbf{B}}^2$ be the estimators of the sliced inverse regression based on these two subsets respectively. We define the following mirror statistics
\begin{equation}\label{eq:mirror}
    M_j = \sum_{h=1}^H \hat{\mathbf{B}}^{1}_{jh} \hat{\mathbf{B}}^{2}_{jh}.
\end{equation}
Note that when the j-th hypothesis is true, namely, $\mathbf{P}_{span(x_j)}( \mathcal{S}_{y|\vx}) = \mathcal{O}_p$, one would expect that both vectors $\hat{\mathbf{B}}_{j\cdot}^1$ and $\hat{\mathbf{B}}_{j\cdot}^2$ would center around zero. %
On the other hand, when $H_j$ is false, both vectors center around a common non-zero vector $\vB_{j\cdot}$. The angles between $\hat{\vB}_{j\cdot}^k, k=1,2,$ and $\vB_{j\cdot}$ are approximately the same. Consequently, we call this statistic $M_j$ a {\bf A}ngular {\bf B}alanced {\bf S}tatistic (ABS). Intuitively, the ABS statistic would be symmetric with respect to zero under the null and it tends to have a positive value under the alternative. %
This is depicted in Figure \ref{fig:demo}. In the left panel, when considering the predictor $x_1 \in \mathcal{MB}(y,\vx)$, the estimated coefficient vectors $\hat{\mathbf{B}}_{1\cdot}^1$ and $\hat{\mathbf{B}}_{1\cdot}^2$ center around a non-zero vector. On the other hand when considering the estimated coefficients of $x_2$ which does not belong to $\mathcal{MB}(y,\vx)$, the estimated coefficients $\hat{\mathbf{B}}_{2\cdot}^1$ and $\hat{\mathbf{B}}_{2\cdot}^2$ center around the origin. In the right panel, we plot the histogram of all the mirror statistics when generating the data according to Setting 1 in Section \ref{sec:simulation}, it is seen that $M_j$'s are roughly symmetric with respect to zero when the corresponding null hypotheses are true. When the corresponding null hypothesis is false, this statistic tends to have a large positive value.

The following theorem provides a rigorous argument for the above intuition. Namely, we show that the ABS is symmetric with respect to zero under the null hypothesis asymptotically. %

\begin{theorem}\label{thm:sym}
Assume the same conditions of Theorem \ref{thm:asym:normal}, if the null hypothesis $\mathcal{H}_j$ is true, then for any $t>0$, with probability at least $1-2\exp(-cn^\alpha)$, 
\begin{equation}\label{eqn:sym}
\sup_t \left|\mathbb{P}\left(  M_j< -t \right) - \mathbb{P}\left(  M_j  > t \right)\right| \le CD_{n,p}.
\end{equation}
\end{theorem}

We now turn to the false discovery rate control. To simplify our notation, let $\theta_j$ be the indicator of whether the $j$-th hypothesis defined in (\ref{eqn:hypothesis}) is false. Let $\cS_0=\{j: \theta_j=0\}$ be the set consisting of all the true hypotheses. Consider the single-step procedure where the $j$-th hypothesis is rejected if the ABS $M_j$ is greater than some threshold $t$. Namely, we choose a decision $\bdelta=(\delta_1,\cdots,\delta_p)$ as $\delta_j=1(M_j>t)$. Then the false discovery proportion is
\[
FDP(t) =\frac{\sum_{j:\theta_j\in \cS_0} \delta_j }{ ( \sum_j \delta_j )\vee 1}.
\]
According to Theorem \ref{thm:sym}, the probabilities $\mathbb{P}(M_j>t)$ and $\mathbb{P}(M_j<-t)$ are approximately the same when the hypothesis $\mathcal{H}_j$ is true. 
Namely,
\begin{align*}
&\mathbb{E}\sum_{j:\theta_j\in\cS_0}\delta_j = \sum_{j:\theta_j\in\cS_0}\mathbb{P}(M_j>t) \\
&\approx \sum_{j:\theta_j\in\cS_0}\mathbb{P}(M_j<-t) \le \sum_j \mathbb{P}(M_j<-t).
\end{align*}

Consequently, we estimate the false discovery proportion by
\begin{equation}\label{eqn:estifdp:mim}
\widehat{FDP(t)} = \frac{ \sum_{j} 1(M_j < -t )}{\sum_j 1(M_j>t)}.
\end{equation}
Note that $\widehat{FDP(t)}$ tends to overestimate the true false discovery proportion. However, this is not a serious issue due to two reasons: (i) the over-estimation protects us from large false positives and (ii) the probability that $\mathbb{P}(M_j<-t)$ is negligible when $\mathcal{H}_j$ is false.

For any designated level, say $q$, we could choose a data-driven threshold $\tau_q$ as
\[
\tau_q = \min \left\{t: \widehat{FDP(t)} \le q\right\}.
\]
We call this method {\bf M}odel free multiple {\bf T}esting procedure based on {\bf A}BS ({\bf MTA}). The steps of MTA are summarized in Algorithm \ref{alg:mmm}.

\begin{algorithm}[ht]
\begin{enumerate}
\item Split the sample as two parts, denoted as $\mathcal{D}_1$ and $\mathcal{D}_2$;

\item For each sample, estimate $\hat{\mathbf{B}}^1$ and $\hat{\mathbf{B}}^2$ according to (\ref{eqn:Bhat});

\item For $j=1, 2,\cdots, p$, calculate the ABS $M_j$ as $M_j = \sum_{h=1}^H \hat{\mathbf{B}}^{1}_{jh} \hat{\mathbf{B}}^{2}_{jh}$;

\item Find the threshold $\tau_q$ such that
\[
\tau_q = \min \left\{t: \widehat{FDP(t)} \le q\right\},
\]
where
\[
\widehat{FDP(t)} = \frac{\sum_{j=1}^p 1 (M_j\le -t)}{\sum_{j=1}^p 1(M_j>t)};
\]
\item For $j=1,2,\cdots, p$, reject $\mathcal{H}_j$ if $M_j > \tau_q$ and fail-to-reject $\mathcal{H}_j$ otherwise.

\end{enumerate}
\caption{Model free multiple Testing procedure based on ABS (MTA)}\label{alg:mmm}
\end{algorithm}

We now turn our attention to the asymptotic property of MTA. We assume the following assumption:

\begin{assumption}\label{assump}
(a) Let $\bSigma^{jj}$ be the covariance matrix of $\vz_{ij\cdot}$, $\bSigma^{j j^\prime}$ be the covariance between $\vz_{ij\cdot}$ and $\vz_{ij^\prime\cdot}$.
We assume that
\begin{equation*}
    \sum_{j,j^\prime \in \cS_k}r_1\left( (\bSigma^{jj})^{-1/2}\bSigma^{jj^\prime}(\bSigma^{j^\prime j^\prime})^{-1} \bSigma^{j^\prime j} (\bSigma^{jj})^{-1/2} \right) < C_1 p_k^{\alpha_1}, 
\end{equation*}
holds for $k=1, 2$, a constant $C_1>0$, and $\alpha_1\in(0,2)$ where $r_1(\cdot)$ denotes the largest singular value.
(b) Let $\rho(z_{ijh}, z_{ij^\prime h})$ be the correlation coefficient of $z_{ijh}$ and $z_{ij^\prime h}$. We assume that
\begin{equation*}
    \sum_{j,j^\prime \in \cS_0}\sum_{h=1}^{H} |\rho(z_{ijh}, z_{ij^\prime h})| < C_2 p_0^{\alpha_2},
\end{equation*}
holds for a constant $C_2>0$ and $\alpha_2\in(0,2)$.
\end{assumption}

For linear model, a similar assumption stating that the sum of partial correlation is bounded by $C_1 p_0^{\alpha}$, where $\alpha\in (0, 2)$ is imposed in \cite{xing2021controlling}. When dealing with nonlinear model, Assumption \ref{assump} requires that the sum of pair-wise partial correlation in each slice being bounded by $C p_0^{\alpha}$ for a constant $C>0$ any $\alpha \in (0,2)$. However, we consider a more general non-linear model where the contribution of the $j$-th variable is related to vectors with length $H$. Assumption 3(a) is equivalent to bound the pairwise canonical correlations which is a general procedure for investigating the relationships between two sets of variables. In Assumption 3(b), we consider the absolute sum of the correlation coefficients of corresponding variables in two sets. Since the correlation measure is not unique, we consider two different types to enlarge the suitable cases for our theory. 
The following two lemmas characterize the sum of pairwise covariances of mirror statistics, which is the key to  establishing the asymptotic property of MTA. 

\begin{lemma}\label{lemma:suff:ols}
Let $M_j$ be the angular balanced statistic defined in (\ref{eq:mirror}). If Assumption~\ref{assump} holds, then with probability at least $1-2\exp(-cn^\alpha)$, 
\begin{eqnarray}\label{def:weak:dep}
&& \sum_{j,k\in \mathcal{S}_0}\cov\left( \mathbbm{1}(M_j\ge t), \mathbbm{1}(M_k\ge t) \right) \le C_1^\prime |\mathcal{S}_0|^{\alpha_1}, \; \forall \;  t,\nonumber
\end{eqnarray}
where $\alpha_1\in(0,2)$, $C_1^\prime>0$ is a constant. %
\end{lemma}
In Lemma \ref{lemma:suff:ols}, we build a relationship between the correlations among mirror statistics and the correlations among covariates. Next, we derive asymptotic properties of the MTA method.  We define a few quantities:
\begin{gather*}
V(t) =\frac{\#\{ j:  j\in \mathcal{S}_0, M_j \le -t \}}{p_0} ,\, V'(t) =\frac{\#\{ j:  j\in \mathcal{S}_0, M_j \ge t \} }{p_0},\, \\
V^{1}(t) =\frac{\#\{ j:  j\in \mathcal{S}_1, M_j \ge t \}}{p_1} , \\
G_0(t) = \lim_p \frac{\sum_{j\in\mathcal{S}_0} \mathbb{E} 1(M_j\le -t)}{p_0},\, G_1(t) = \lim_p \frac{\sum_{j\in\mathcal{S}_1} \mathbb{E} 1(M_j\ge t)}{p_1},
\end{gather*}
and
\begin{align*}
    &FDP(t) :=\frac{V^{\prime}(t)}{(V^{\prime}(t) + r_p V^1(t))\vee 1/p},\\
    &FDP^{\infty}(t) :=\lim_{p}\frac{V^{\prime}(t)}{(V^{\prime}(t) + r_p V^1(t))\vee 1/p},
\end{align*}
where $r_p = p_0/ p_1$ and $FDP^{\infty}(t)$ is the pointwise limit of $FDP(t)$.  We have the following results.

\begin{lemma}\label{lem:slln}
Suppose Assumption \ref{assump}(a) or (b) holds and $G_0(t)$ is a continuous function. Then, we have
\begin{align*}
& \sup_t \left|{V(t)} - G_0(t)\right|\xrightarrow[]{p}0,\quad  \sup_t\left|{V'(t)}- G_0(t)\right|\xrightarrow[]{p}0,\\
&\sup_t\left|{V^1(t)}- G_1(t)\right|\xrightarrow[]{p}0.
\end{align*}
\end{lemma}

Lemma 2 is based on the weak dependence assumption 3(a) or 3(b) under a model-free assumption. In literature, similar week dependence conditions on p-values \citep{storey2004strong} and linear models \citep{dai2020false,xing2021controlling, dai2022false} are commonly used to study the asymptotical property of FDR. With the aid of this lemma, we show that the FDR of the proposed MTA method is less than or equal to $q$ asymptotically.  
\begin{theorem}\label{thm:fdr}
For any given level $q \in (0,1)$, assume that the pointwise limit $\text{FDP}^\infty(t)$ of $\text{FDP}(t)$ exists for all $t > 0$, and there is a constant $t_q > 0$ such that $\text{FDP}^\infty(t_q) \leq q.$ When Assumption (\ref{assump:1}-\ref{assump}) holds and $\mathcal{P}_H\in \mathcal{DP}_H(b)$, we have
\begin{equation}\nonumber
\text{FDP} \leq q + o_p(1)\ \ \ \text{and}\ \ \ \limsup_{n,p\to\infty}\text{FDR} \leq q.
\end{equation}
\end{theorem}

The proof of Theorem \ref{thm:fdr} is included in the appendix. %

\section{Numerical Studies}\label{sec:simulation}
In this section, we use simulations to evaluate the finite sample performance of MTA with state-of-the-art competitors. First, we consider the model-X knockoff, which constructs knockoffs $\tilde{\vx}$ by assuming the distribution of $\vx$ is Gaussian with a known covariance matrix. We construct the feature statistic as
\begin{equation*}
    W_j =  \sum_{h=1}^H \hat{\mathbf{B}}_{jh} \hat{\mathbf{B}}_{\tilde{j}h},
\end{equation*}
where $\hat{\mathbf{B}}$ is the solution of (\ref{eqn:Bhat}) when replacing $\vx$ by $(\vx, \tilde{\vx})$.
Second, we consider the marginal non-parametric test to test the dependence between the response variable and each predictor and calculate the Hilbert-Schmidt independence criterion (HSIC) test statistic proposed in  \cite{gretton2007kernel}. After obtaining the p-values, we apply the Benjamini-Hochberg (BH) procedure (HSIC+BH). Third, we apply the dimension reduction coordinate test (DR-coor) in \citet{cook2004testing} to test the contribution of each predictor and calculate the p-value followed by the BH method (DR-coor+BH). In the following sections, we consider three different multi-index models with a wide range of nonlinear dependencies between the response and predictors.
For predictors, we consider two different designs including Gaussian design  and a real-data design.

\subsection{Simulation with Gaussian Design}
We set $n=1000$, $p=200$. The covariates are generated from multivariate Gaussian distribution $N(0, \Sigma)$. The covariance matrix $\Sigma$ is autoregressive, i.e., $\sigma_{ij}$, the element at the i-th row and j-th column is $ \rho^{|i-j|}$, where $\rho$ is taken among $0, 0.2, 0.4, 0.6,$ and $0.8$, respectively. We randomly set $20$ nonzero coefficients for each indices and generate the non-zero coefficients from $N(0, 20/\sqrt{n})$. 

Setting 1: 
\begin{equation*}
 y = f_1(\vx)+ \sigma\epsilon = \sin(\va_1^T \vx ) + (\va_2^T \vx)^3 + \sigma\epsilon,
\end{equation*}

Setting 2:
\begin{equation*}
 y = f_2(\vx) + \sigma\epsilon = \frac{(3\va_1^T \vx )}{ 0.5+ (1.5+\va_2^T \vx)^2} + \sigma\epsilon,
\end{equation*}

Setting 3:
\begin{align*}
 y = f_3(x, \epsilon)  =  (a_1^T x)^3 + (a_2^T x ) \times \sigma\epsilon,
\label{eq:model-ex1}
\end{align*}

Setting 4:
\begin{equation*}
 y = f_4(\vx, \epsilon)= \frac{1}{1+ \exp{[\sum_{\ell=1}^{5} \rm{Relu}(\va_\ell^T \vx) } ]} + \sigma\epsilon,
\end{equation*}
where $\rm{Relu}(\cdot) = \max(0, \cdot)$ is a popular choice of activation function in the context of artificial neural networks. This setting is equivalent to a fully connected neural network with one hidden layer.

In the settings 1 to 4, $\epsilon$ is generated using a standard Gaussian distribution. The value of $\sigma_1$ is kept at 0.5 for the first experiment. For $j= 1,2,3$, we adjust $\sigma_j$ so that the ratio $\var(f_j(x))/\sigma_j$ is equal to $\var(f_1(x))/\sigma_1$. This means that the signal-to-noise ratios are kept constant across all the experiments. For the first three experiments, we use two indexes, while in the fourth experiment, we increase the number of indexes to five. Moreover, we treat the error as homoscedastic in experiments 1, 2, and 4. However, in the third experiment, we account for the error as heteroscedastic

Note that the Markov blanket is $\mathcal{MB}(y,\vx) = \{x_j| a_{1j}\ne 0 \quad \mbox{or} \quad a_{2j}\ne 0 \}$.
To evaluate the performance, we calculate the number of true positives (TPs) as the number of variables in $\mathcal{MB}(y,\vx)$ which is selected by a particular procedure, and the number of false positives (FPs) as the difference between the total number of selected variables and TPs. The empirical power is defined as $\#\{TPs\}/ \#\{\mathcal{MB}(y,x)\}$ and the empirical FDP is defined as $\#\{FPs\}/ \#\{\mbox{selected variables}\}$. In all the settings, we set the number of slices $H$ as $20$ for our proposed method.

\begin{figure*}[ht!]
\centering 
\begin{tabular}{cc}
   {\hspace{-30pt}\footnotesize (a1) Power for setting 1}& {\hspace{-30pt}\footnotesize (a2) FDP for setting 1} \\
    \includegraphics[width=0.45\textwidth]{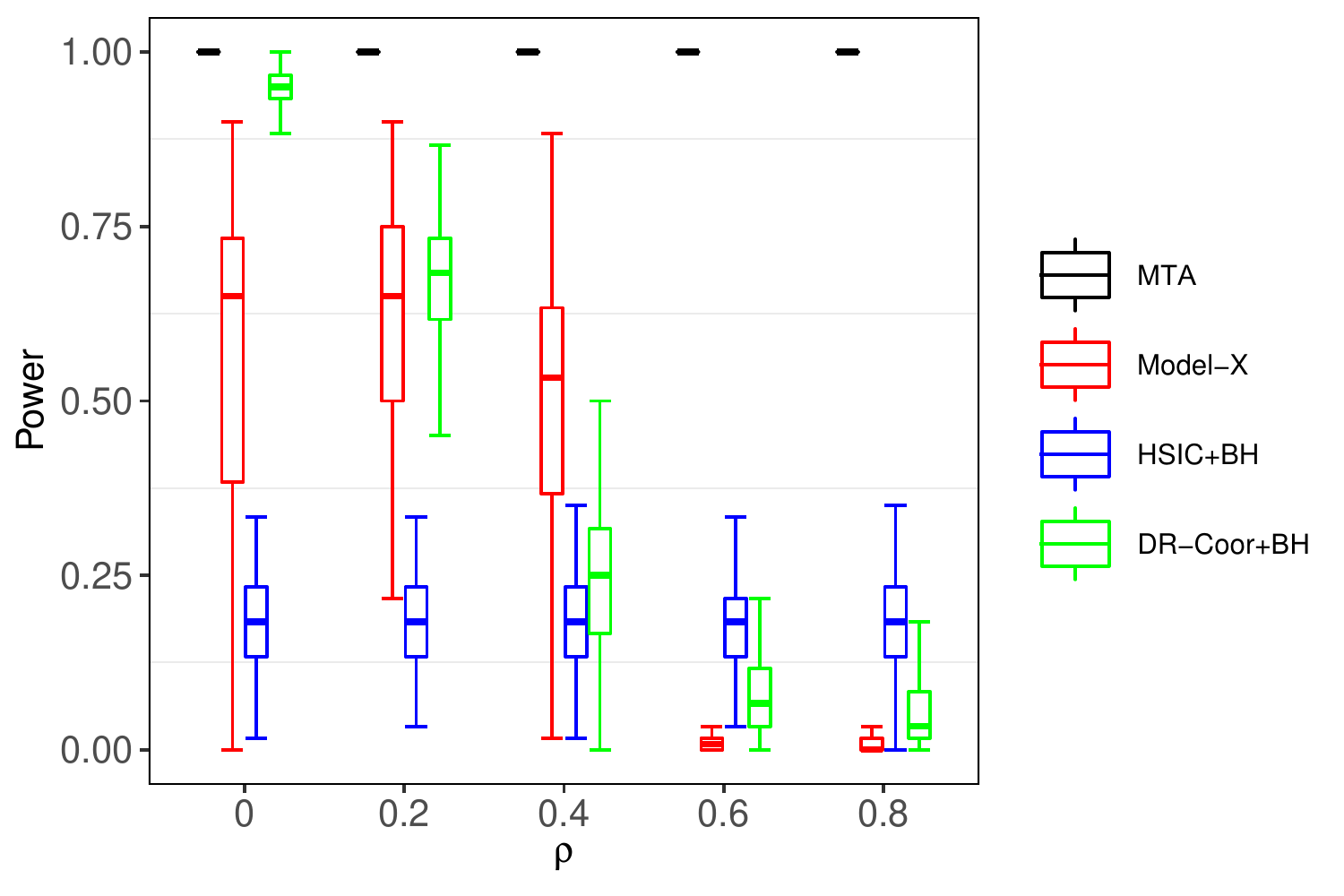}&\includegraphics[width=0.45\textwidth]{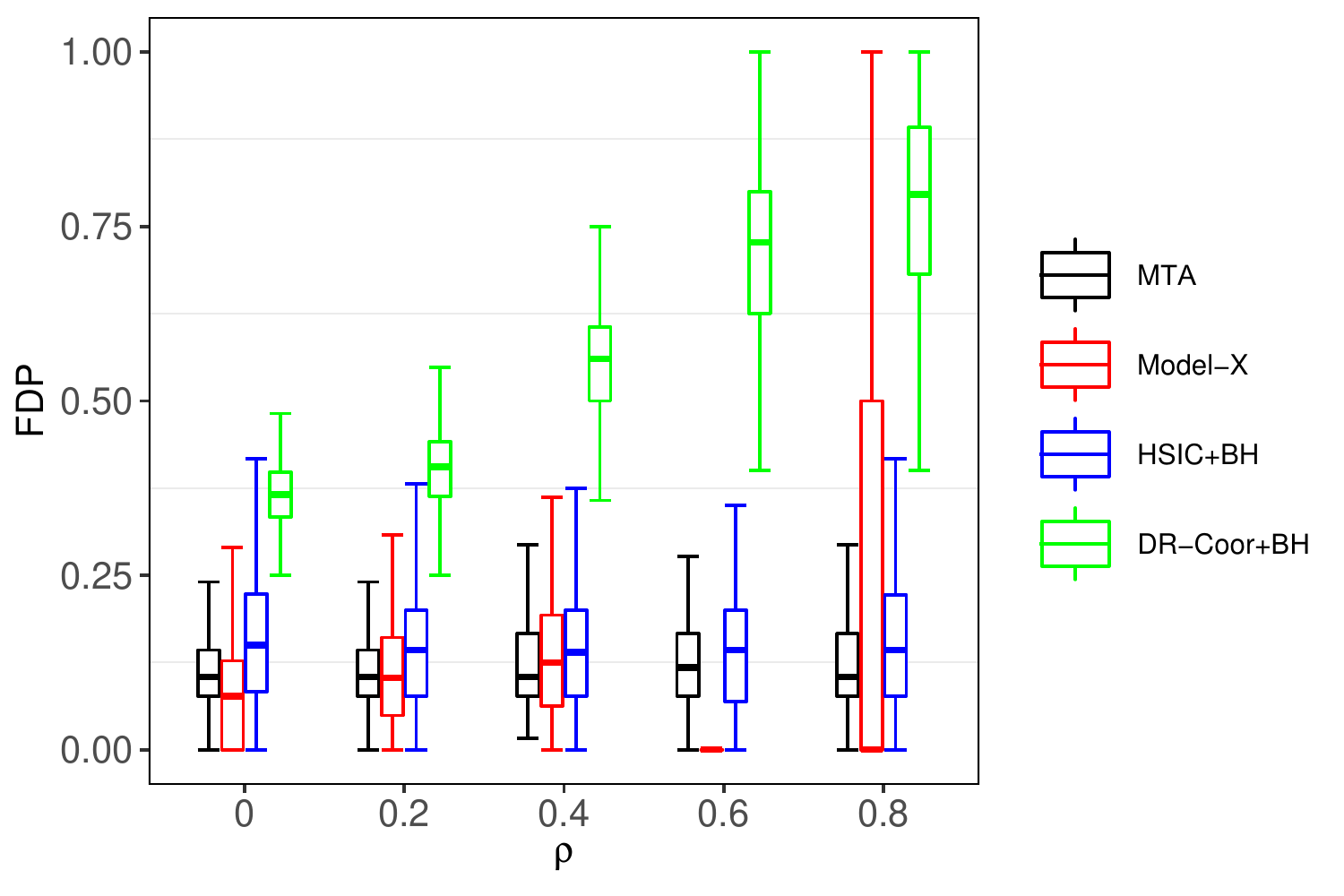} \\ 
    {\hspace{-30pt}\footnotesize (b1) Power for setting 2 }& {\hspace{-30pt}\footnotesize (b2) FDP for setting 2 } \\
    \includegraphics[width=0.45\textwidth]{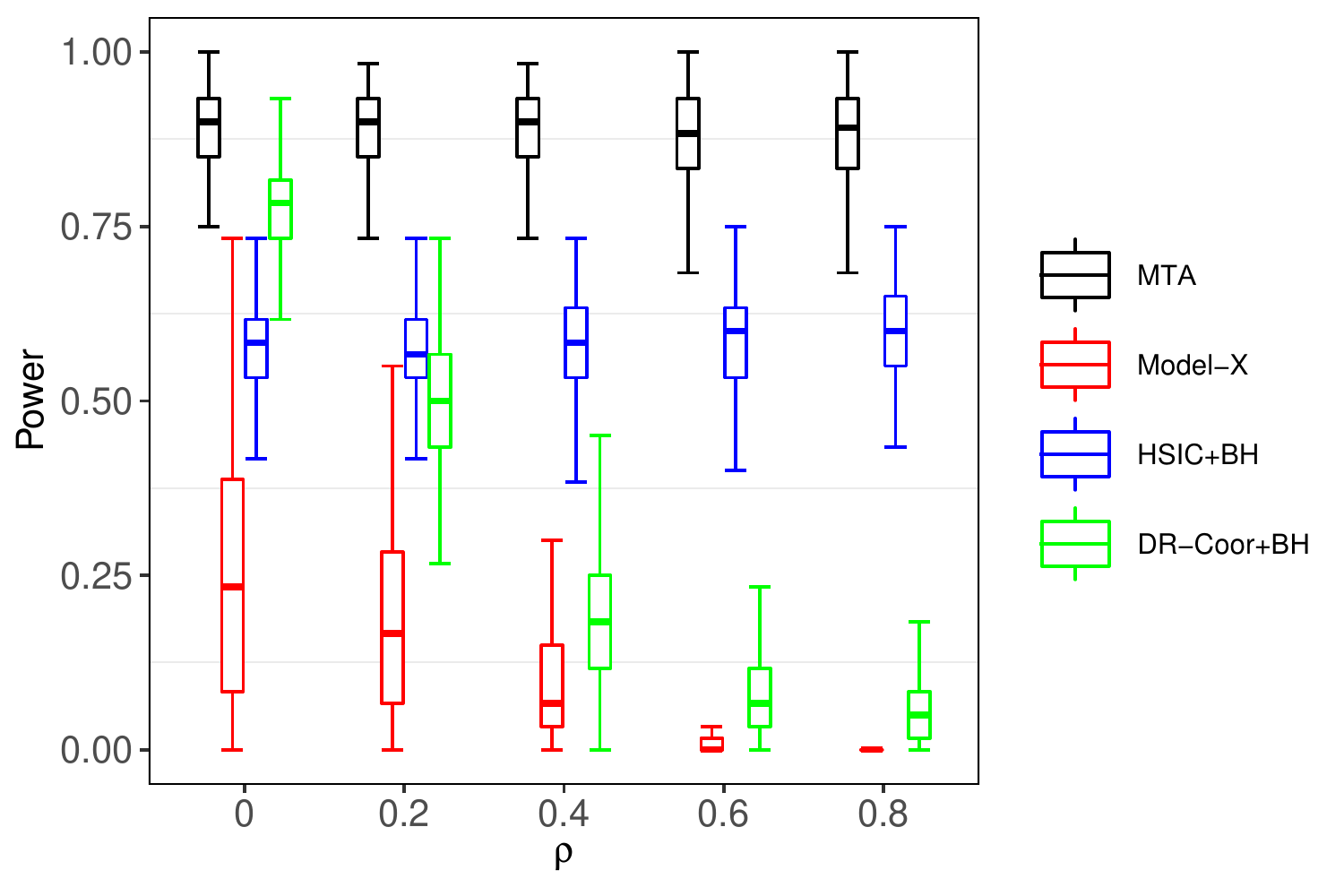}&\includegraphics[width=0.45\textwidth]{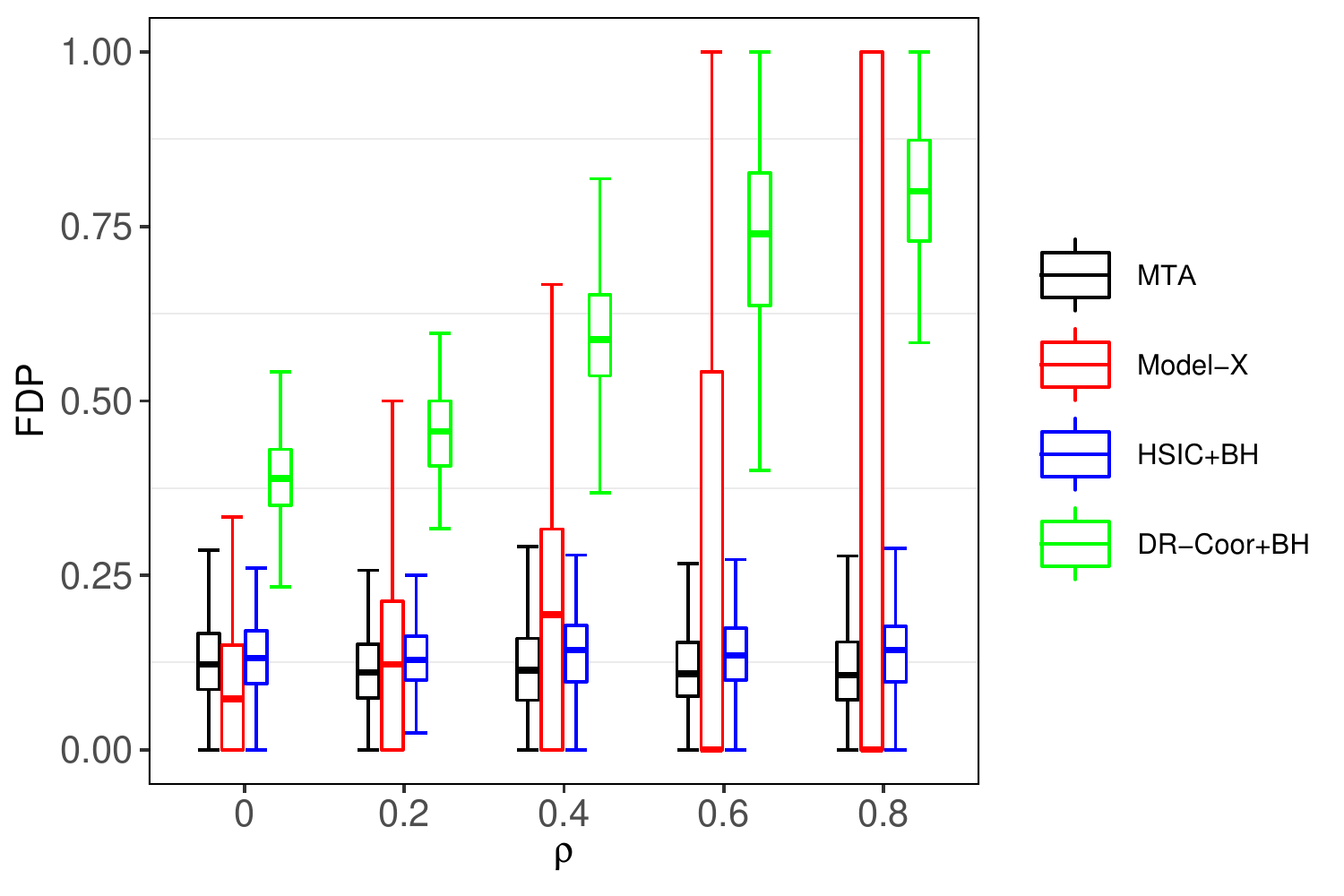}\\
   {\hspace{-30pt}\footnotesize (c1) Power for setting 3 }& {\hspace{-30pt}\footnotesize (c2) FDP for setting 3 } \\
    \includegraphics[width=0.45\textwidth]{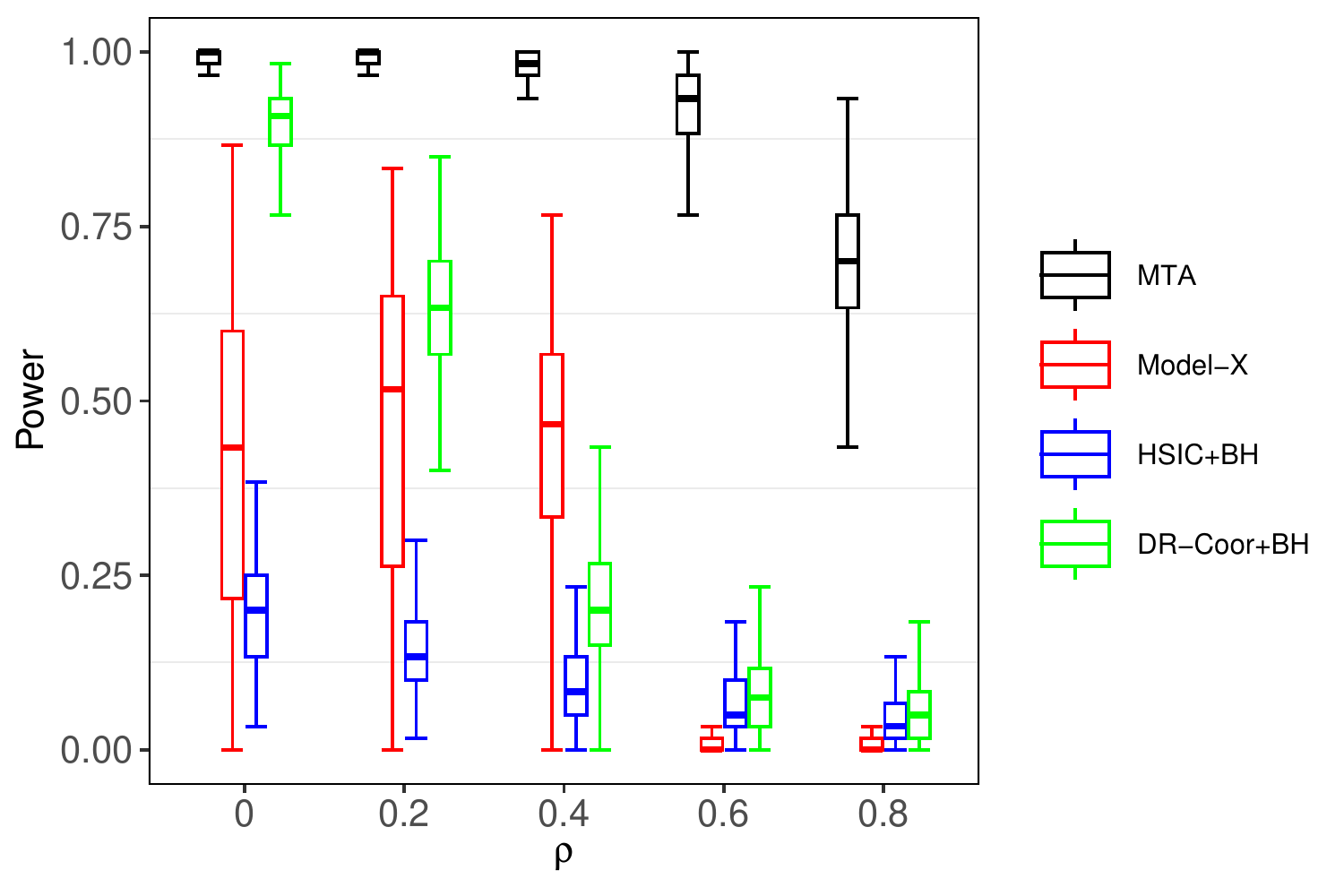}&\includegraphics[width=0.45\textwidth]{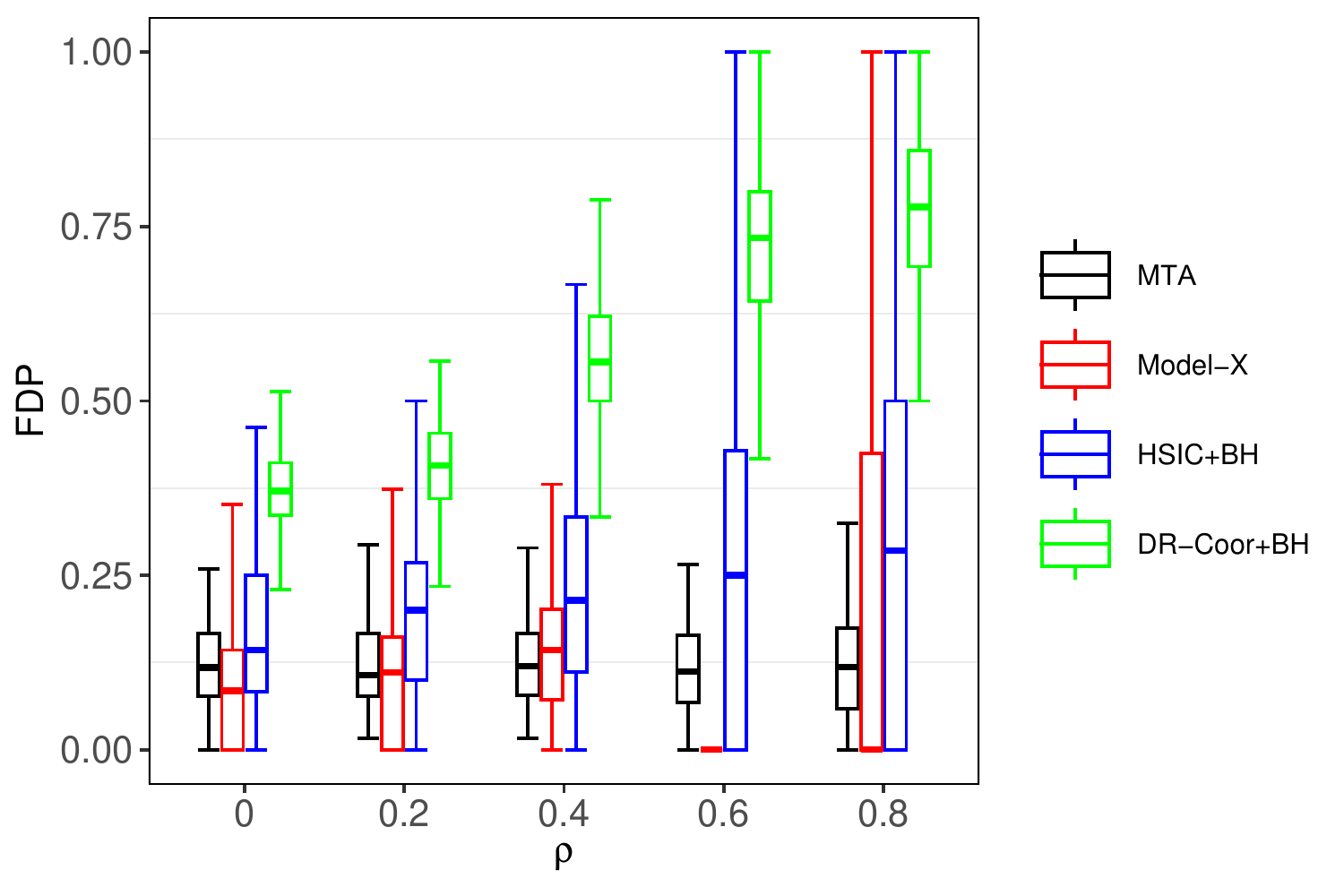}\\
    {\hspace{-30pt}\footnotesize (d1) Power for setting 4 }& {\hspace{-30pt}\footnotesize (d2) FDP for setting 4 } \\
    \includegraphics[width=0.45\textwidth]{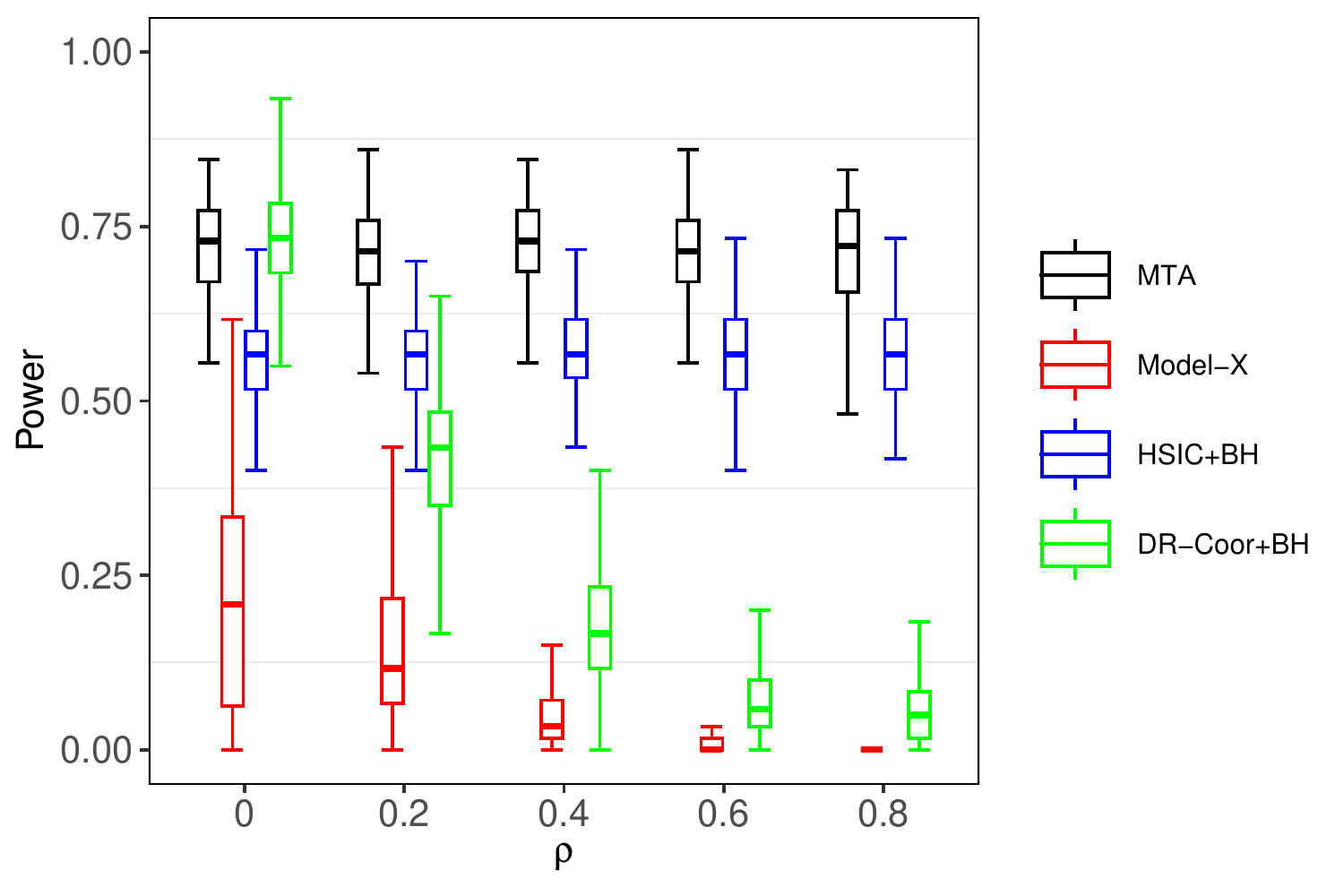}&\includegraphics[width=0.45\textwidth]{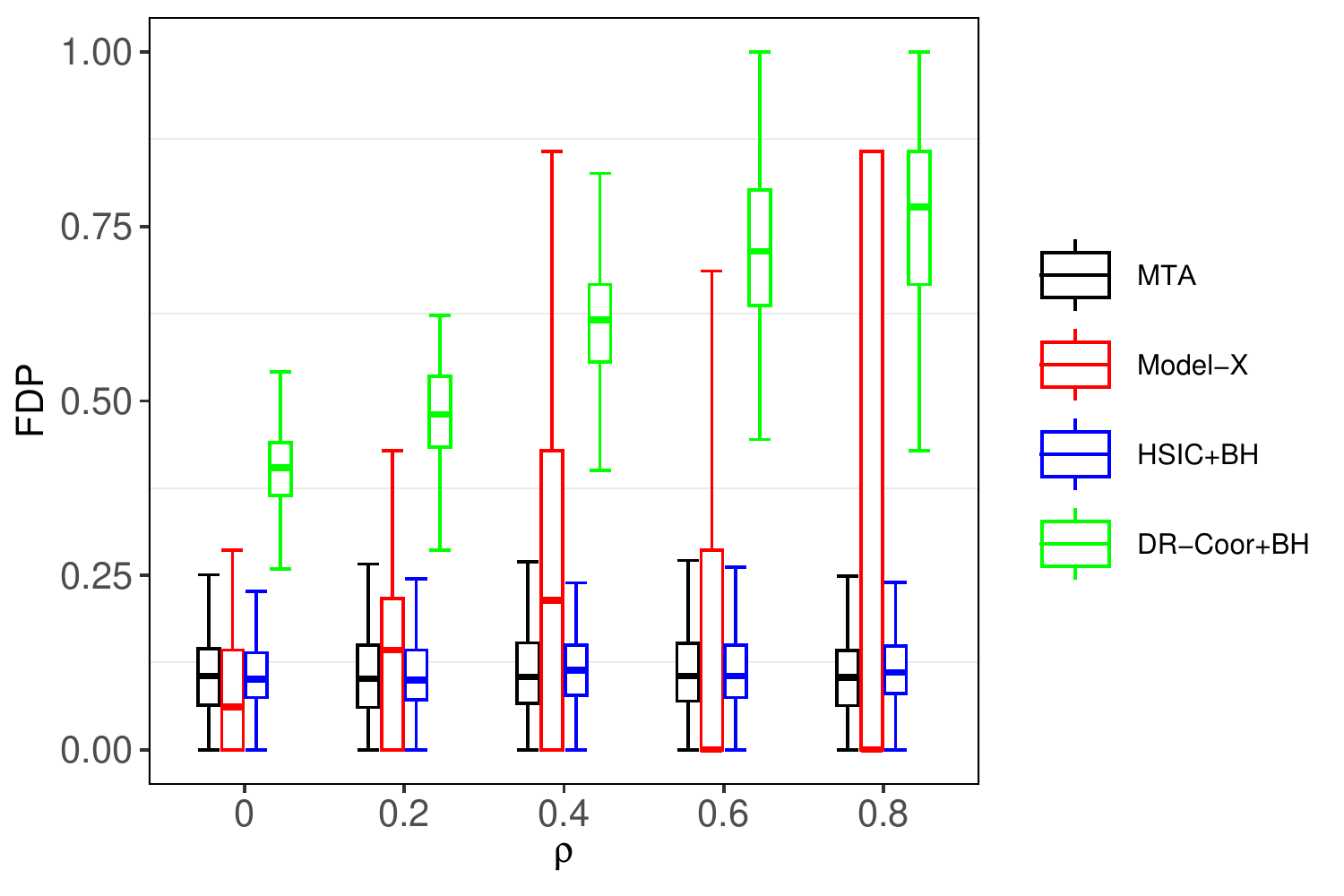}
\end{tabular}
 \caption{Empirical power and FDP for setting 1-3. Here the design matrix is generated from the multivariate normal distribution. }\label{fig:sim-1}
\end{figure*}

In Figure \ref{fig:sim-1}, we plot the boxplot of simulated FDP and power of the considered methods under setting 1-4. 
It is seen that FDR of MTA is controlled at the predefined level under all settings and MTA has larger power than its competitors. We observed that FDPs obtained by Model-X knockoff tend to be either conservative or exceed the predefined levels by a large margin when the correlation is high. This phenomenon has also been observed in \cite{xing2021controlling, candes2018panning}. HSIC-BH methods control the FDR well in settings 1, 2 and 4. But it has a significantly inflated FDR in setting 3 where the noise is heteroscedastic. The power of HSIC-BH methods is lower than our proposed method. 
The DR-Coor-BH test achieves high power when $\rho =0$. However, its power decreases significantly when $\rho$ increases. The FDR is inflated when $\rho$ increases.

\subsection{Simulation studies based on brain connectome}

In this section, we consider using the real design matrix extracted from brain connectome data in the human connectome project (HCP) dataset.
HCP dataset aims at characterizing human brain connectivity in about 1,200 healthy adults to enable detailed comparisons between brain circuits, behavior, and genetics at the level of individual subjects. Customized scanners were used to produce high-quality and consistent data to measure brain connectivity. The data containing various traits and MRI data can be easily accessed through (db.humanconnectome.org).
The real design matrix includes variables with a more complex correlation structure. Moreover, the marginal distributions of variables include skewed, heavy-tailed, and multi-modal distributions. To be consistent with our Gaussian setting, we randomly select $p=200$ brain connections as our predictors and synthetic $y$ using $f_1, f_2, f_3$ listed in our settings 1-3. In this case, we consider different signal to noise level by generating the noise with $\sigma = \psi/\sqrt{n}$ where $\psi=10,15,20,25$.

\begin{figure*}
\centering 
\begin{tabular}{cc}
   {\hspace{-30pt}\footnotesize (a1) Power for setting 1}& {\hspace{-30pt}\footnotesize (a2) FDP for setting 1} \\ 
    \includegraphics[width=0.45\textwidth]{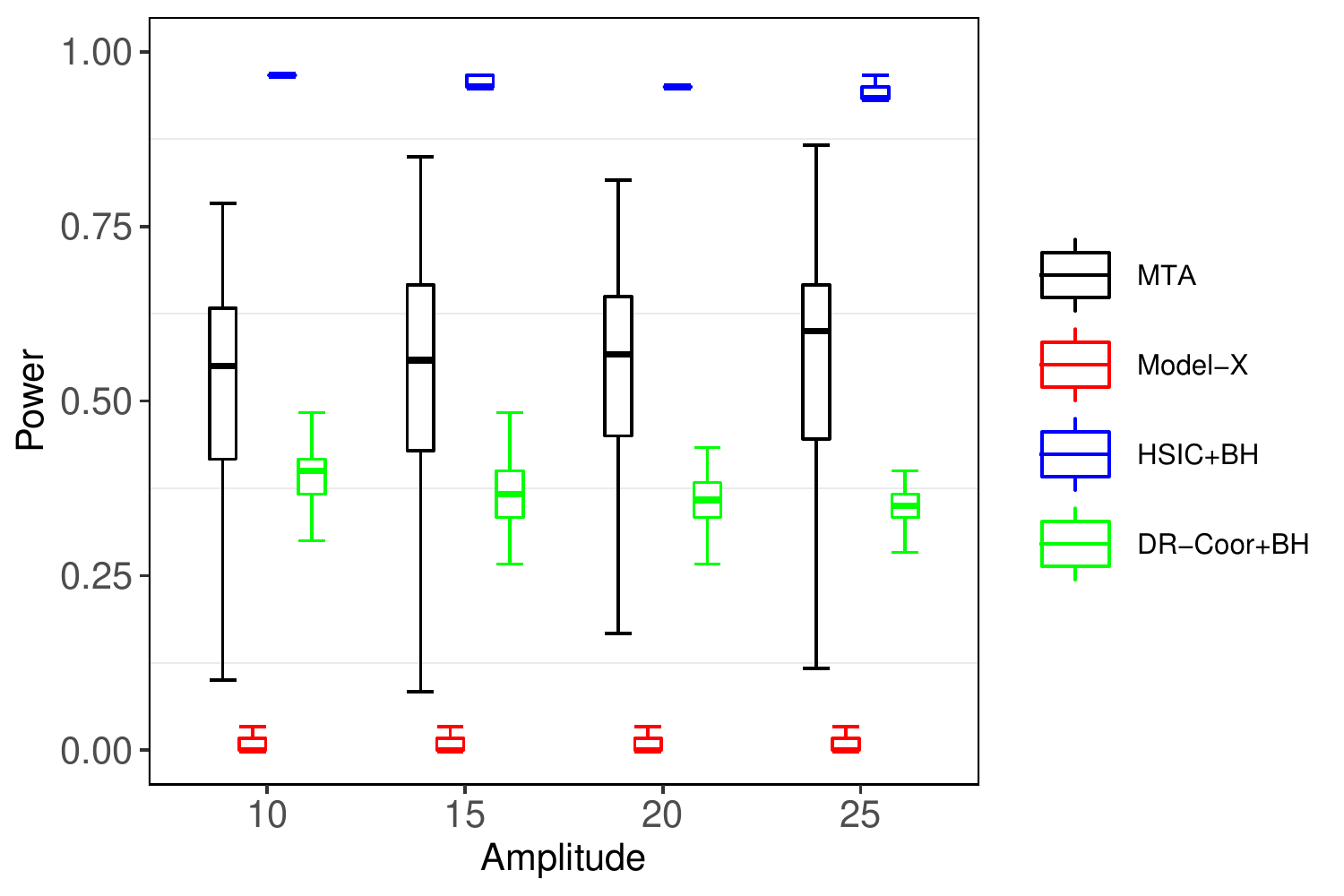}&\includegraphics[width=0.45\textwidth]{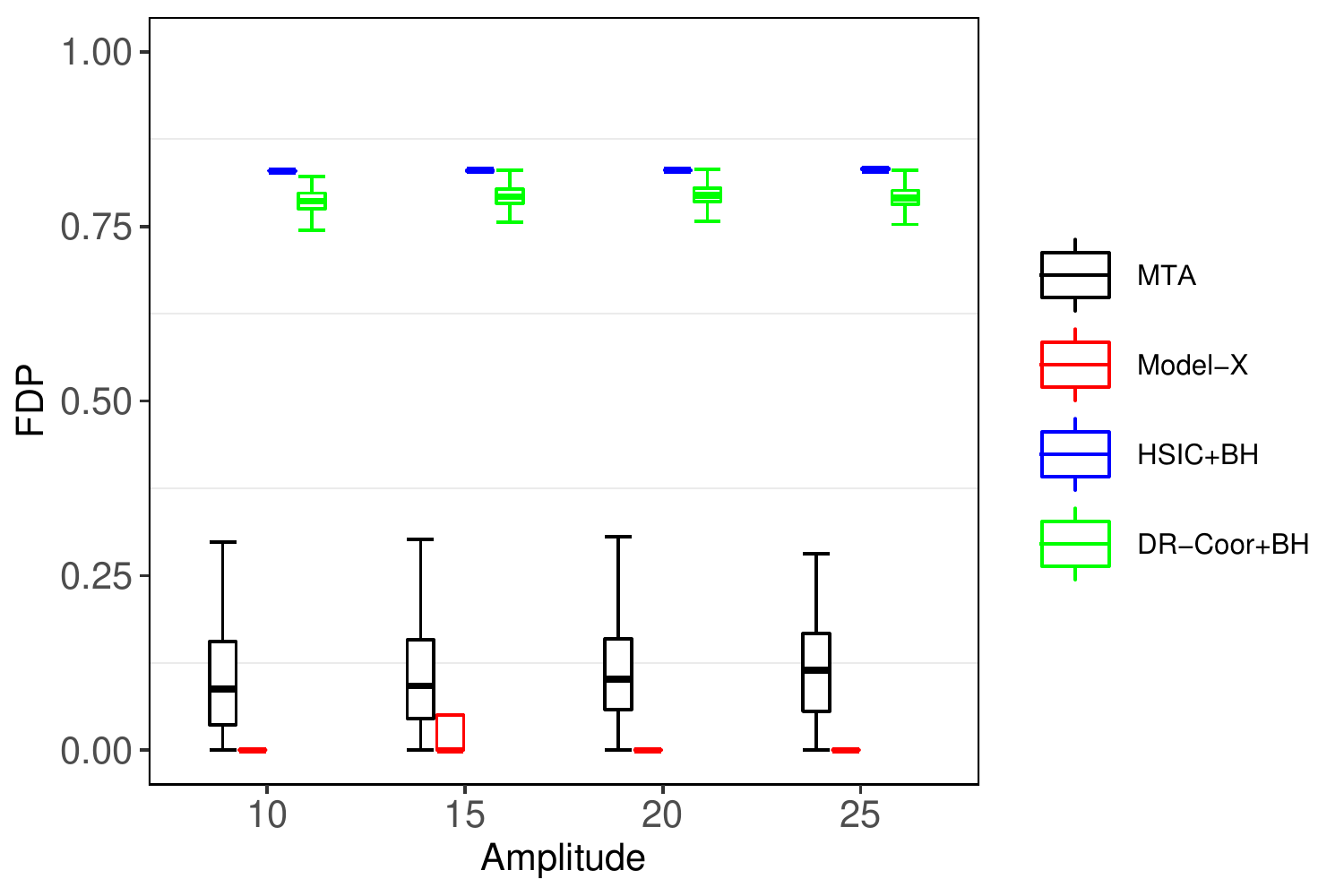} \\ 
    {\hspace{-30pt}\footnotesize (b1) Power for setting 2 }& {\hspace{-30pt}\footnotesize (b2) FDP for setting 2 } \\
    \includegraphics[width=0.45\textwidth]{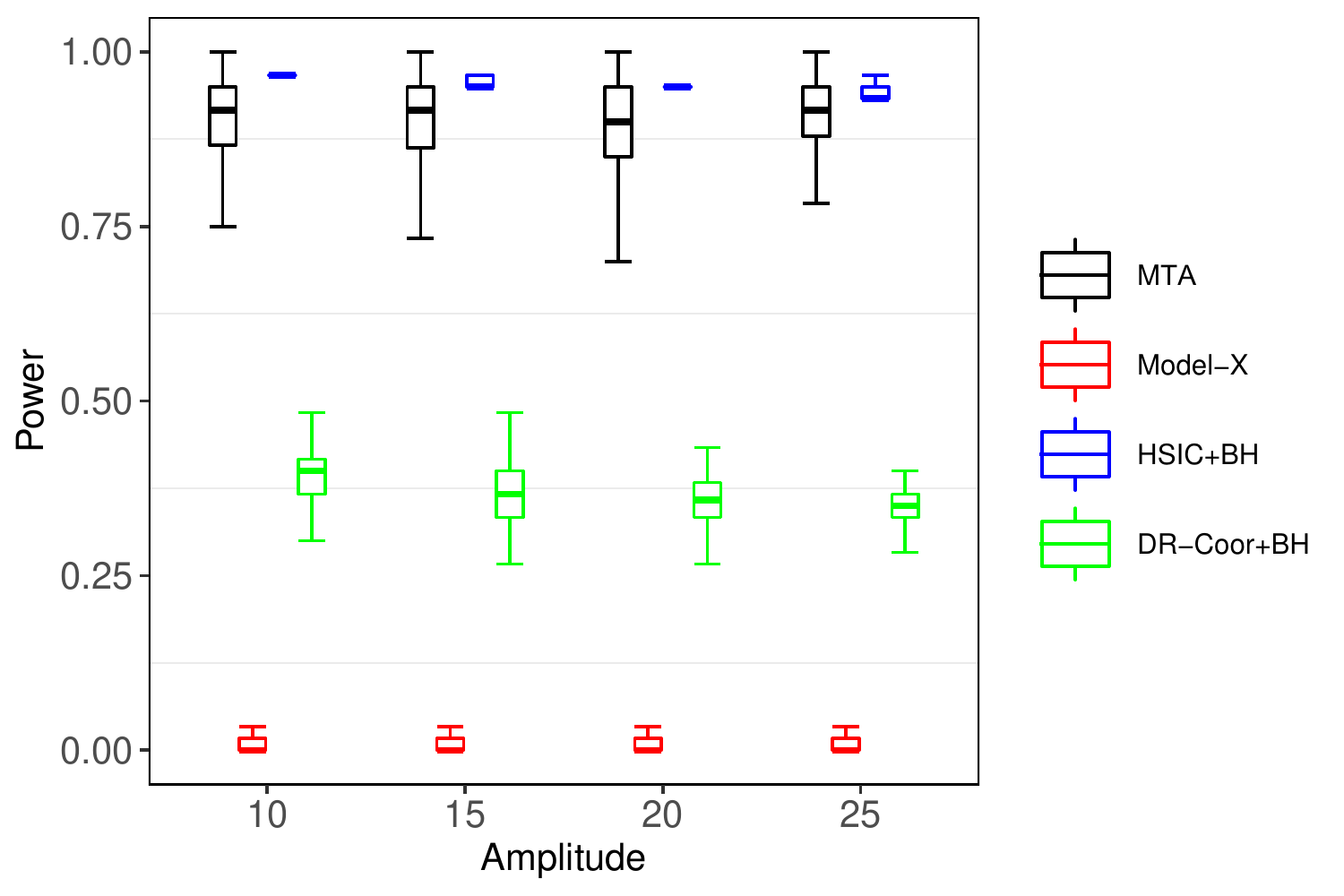}&\includegraphics[width=0.45\textwidth]{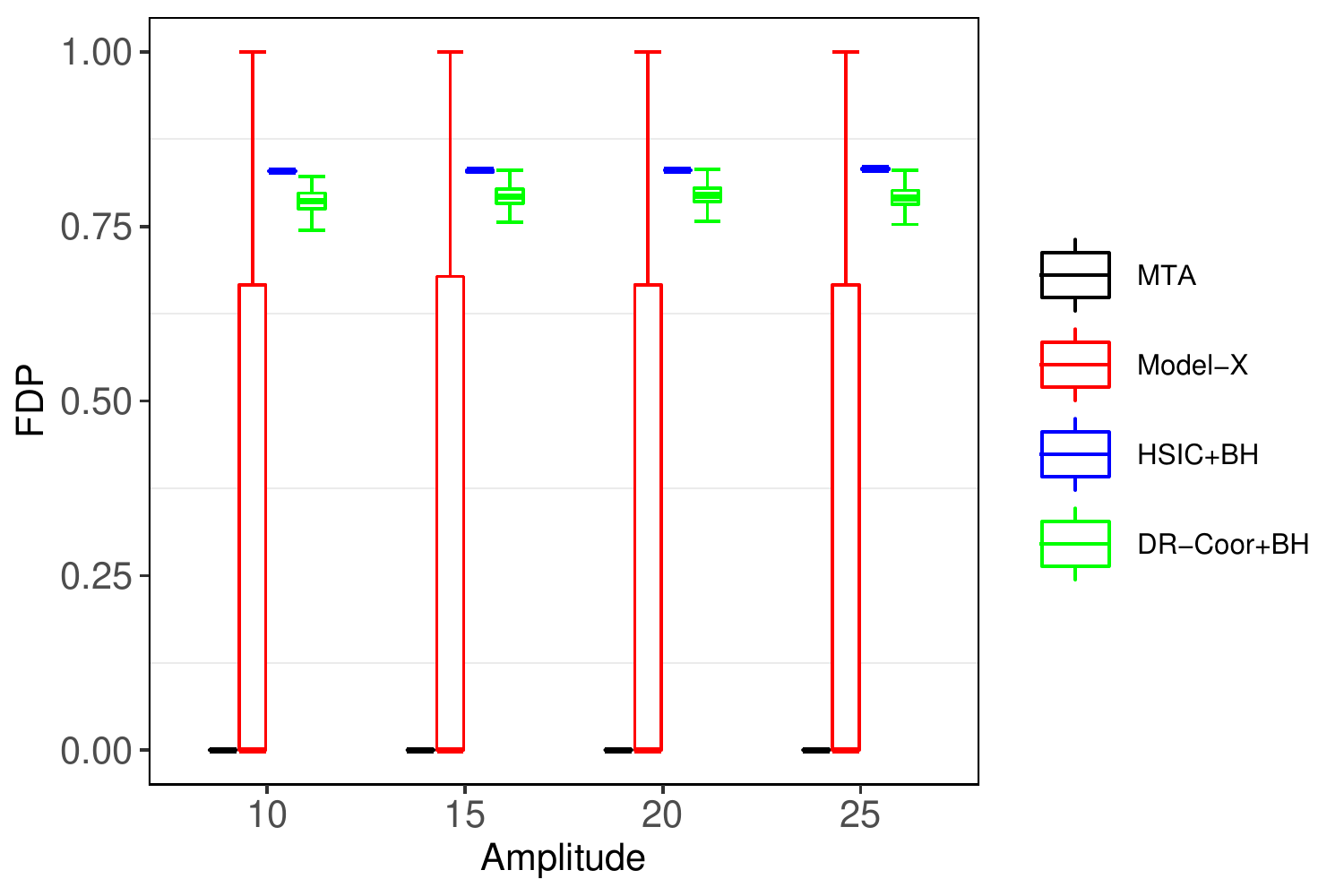}\\
        {\hspace{-30pt}\footnotesize (c1) Power for setting 3 }& {\hspace{-30pt}\footnotesize (c2) FDP for setting 3 } \\
    \includegraphics[width=0.45\textwidth]{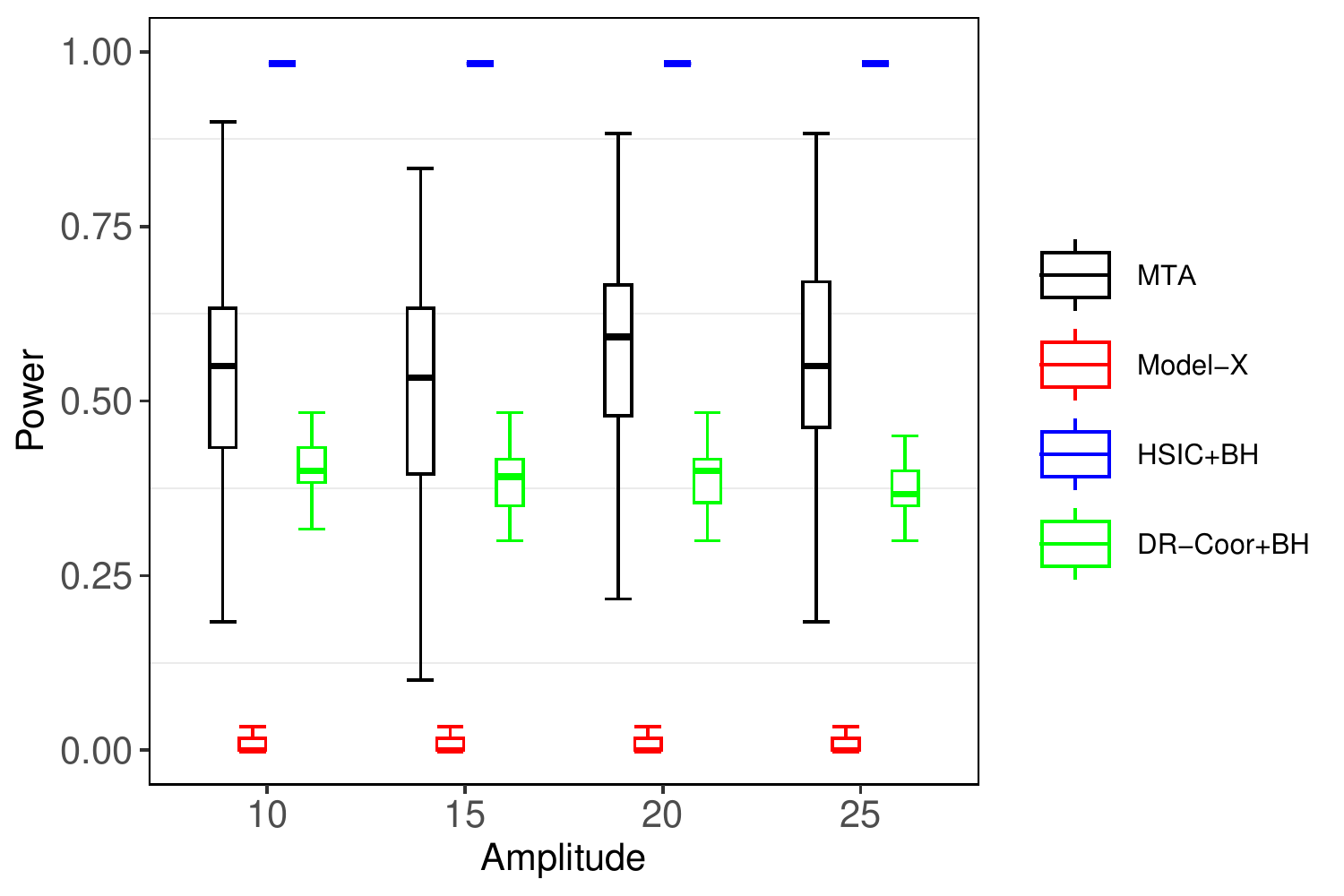}&\includegraphics[width=0.45\textwidth]{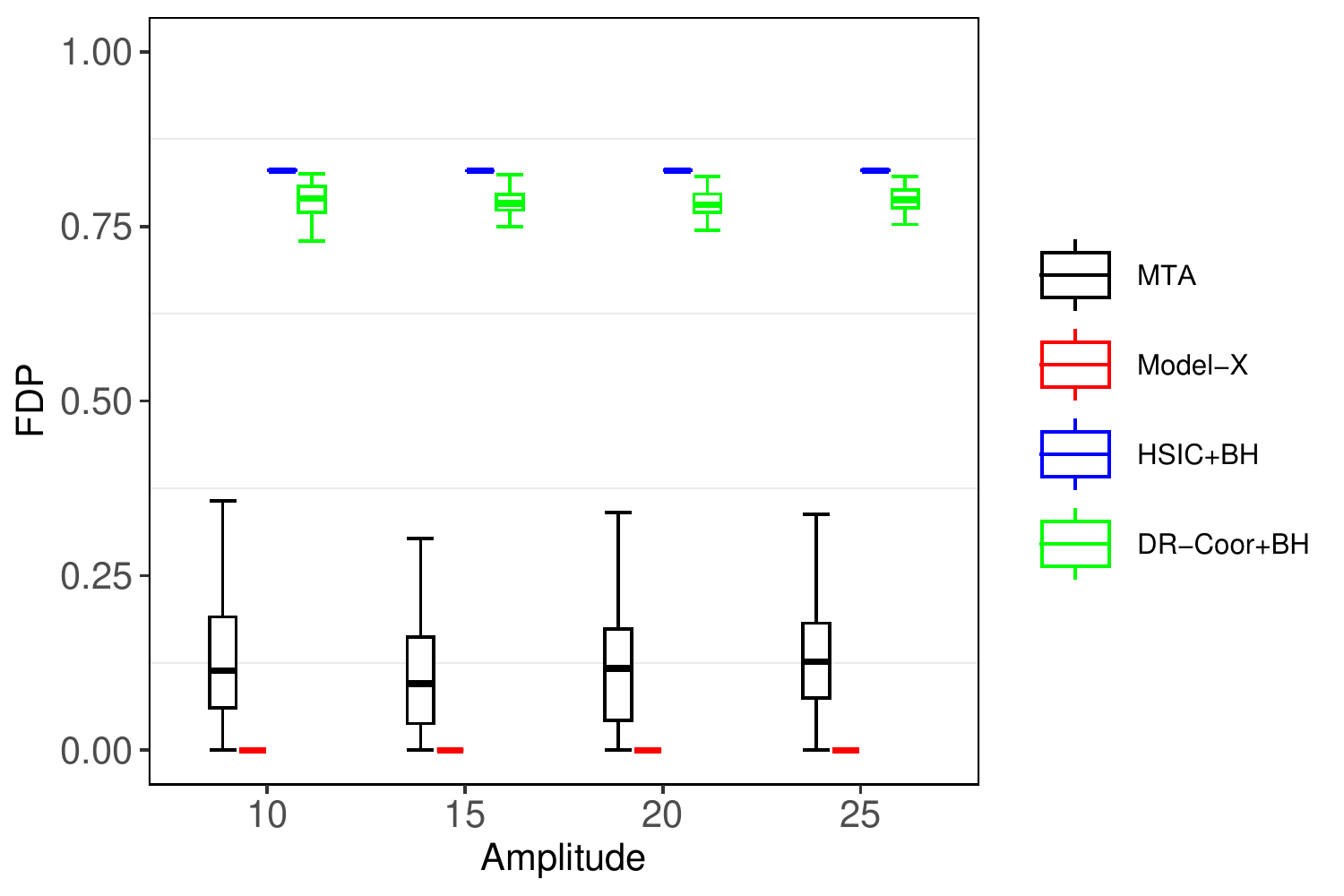}\\
    {\hspace{-30pt}\footnotesize (d1) Power for setting 4 }& {\hspace{-30pt}\footnotesize (d2) FDP for setting 4 } \\
    \includegraphics[width=0.45\textwidth]{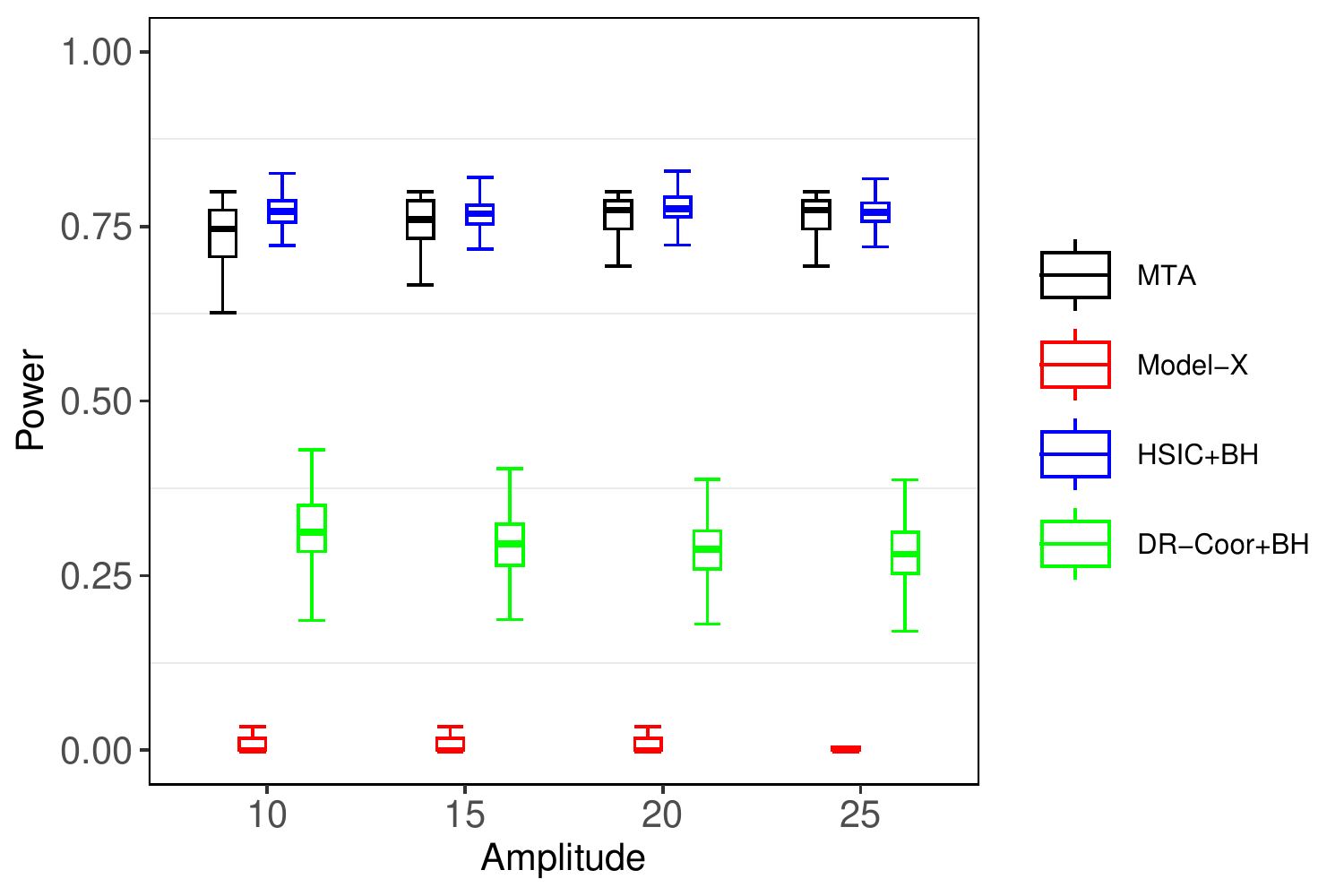}&\includegraphics[width=0.45\textwidth]{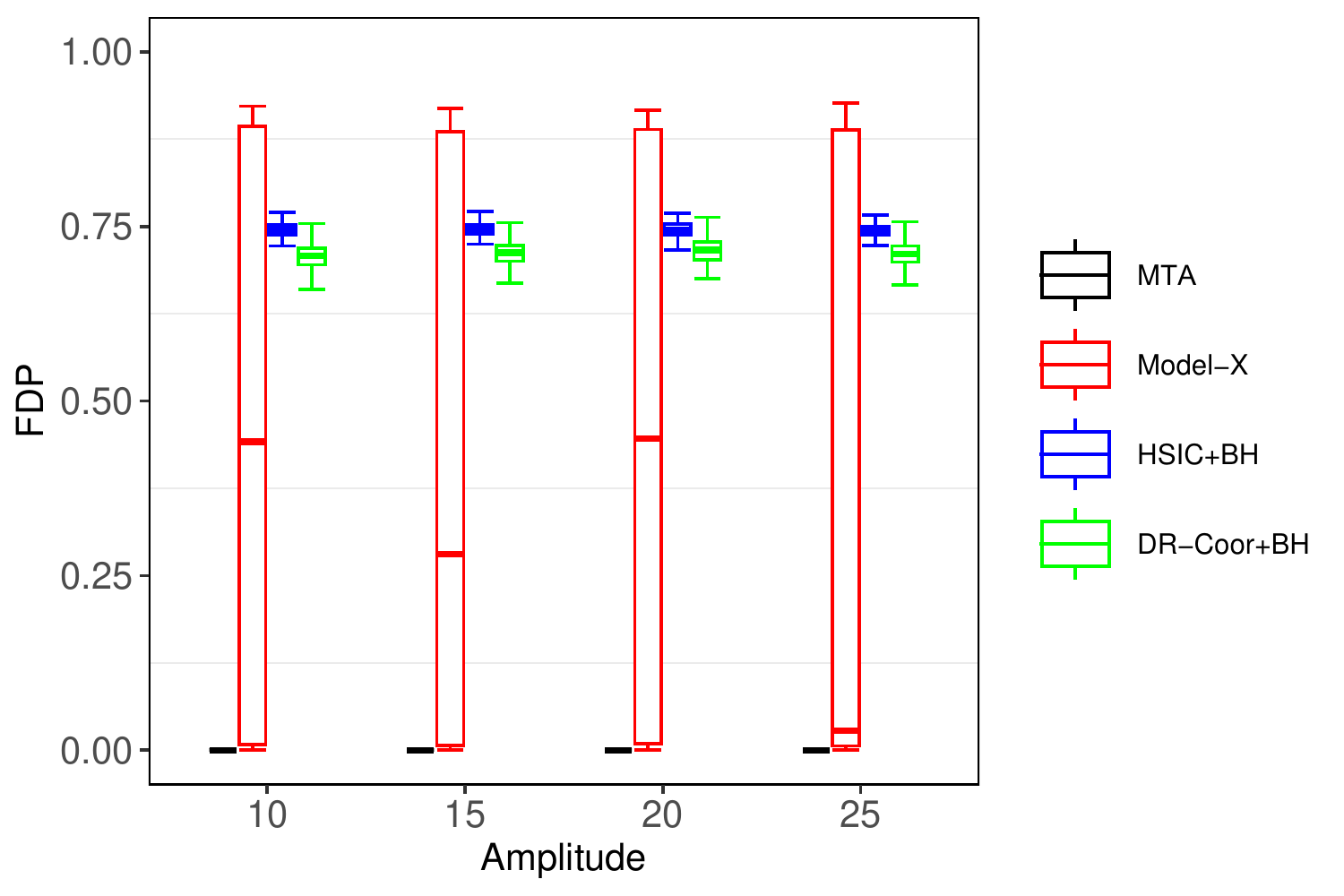}
\end{tabular}
 \caption{Empirical power and FDP for settings 1-3. Here the design matrix is generated from the brain connectome data. }\label{fig:simhcp}
\end{figure*}

It is seen that the proposed MTA method is robust under complex correlation structures. As shown in Figure \ref{fig:simhcp}, for settings 1-4, the MTA method maintains high power subject to the control of FDR under different signal-to-noise levels. Since the distribution of brain connectomes is highly non-Gaussian, the power of the Model-X knockoff method becomes conservative with diminishing power. In contrast, HIC+BH and DR-Coor+BH can not control the FDR well partially due to the high correlation among the covariance in brain connectome data.

\section{Real Data Analysis on Brain Connectome}\label{sec:realdata}

Neural imaging studies hold great promise for predicting and reducing psychiatric disease burden and advancing our understanding of the cognitive abilities that underlie humanity’s intellectual feats. Functional magnetic resonance imaging (fMRI) data are increasingly being used for the ambitious task of relating
individual differences in studying brain function to typical variations in complex psychological phenotypes. In functional brain imaging research, e.g., via fMRI, functional network analysis, which focuses on understanding the connections between different brain areas using some quantitative methods, has become a booming area popular approach. 
The human connectome project (HCP) dataset aims at characterizing human brain connectivity in about 1,200 healthy adults to enable detailed studies of the association between brain functions and behaviors at the level of individual subjects. Customized scanners were used to produce high-quality and consistent data to measure brain connectivity. The data containing various traits and fMRI data can be easily accessed through (db.humanconnectome.org)

To obtain functional connectomes, we used a state-of-the-art fMRI data preprocessing framework – population-based functional connectome (PSC) mapping. The dataset used in this work comes from the HCP Q1 release \citep{van2013wu} consisting of fMRI scans from 68 subjects during resting state (rfMRI). The acquisition parameters are as follows: 90×104 matrix, 220mm FOV, 72 slices, TR=0.72s, TE=33.1ms, flip angle=52°. Preprocessing steps of the rfMRI data have been conducted within the Q1 release, including motion correction, spatial smoothing, temporal pre-whitening, slice time correction, and global drift removal. The fMRI data from each individual in each task was then linearly registered to the MNI152 template.

Since we are
particularly interested in cognition, we extract four cognition-related measures as y from HCP, including:
\begin{itemize}[leftmargin=*]
\item Oral reading recognition test: participants on this test are asked to read and pronounce letters

\item Crystallized composite score: crystallized cognition composite can be interpreted as a global
assessment of verbal reasoning. We use the age-corrected standard score.
\end{itemize}
Our predictors are brain functional connections between each pair of ROIs. Based on the brain parcellation, we have $68*68/2$ connections. We first applied a screening procedure to select top $1/5$ of the connections as our candidates.
We applied the proposed MTA method to select important connections associated with the cognitive traits with controlled false discovery rate at $10\%$.

\begin{figure*}[h]
  \includegraphics[width=\textwidth]{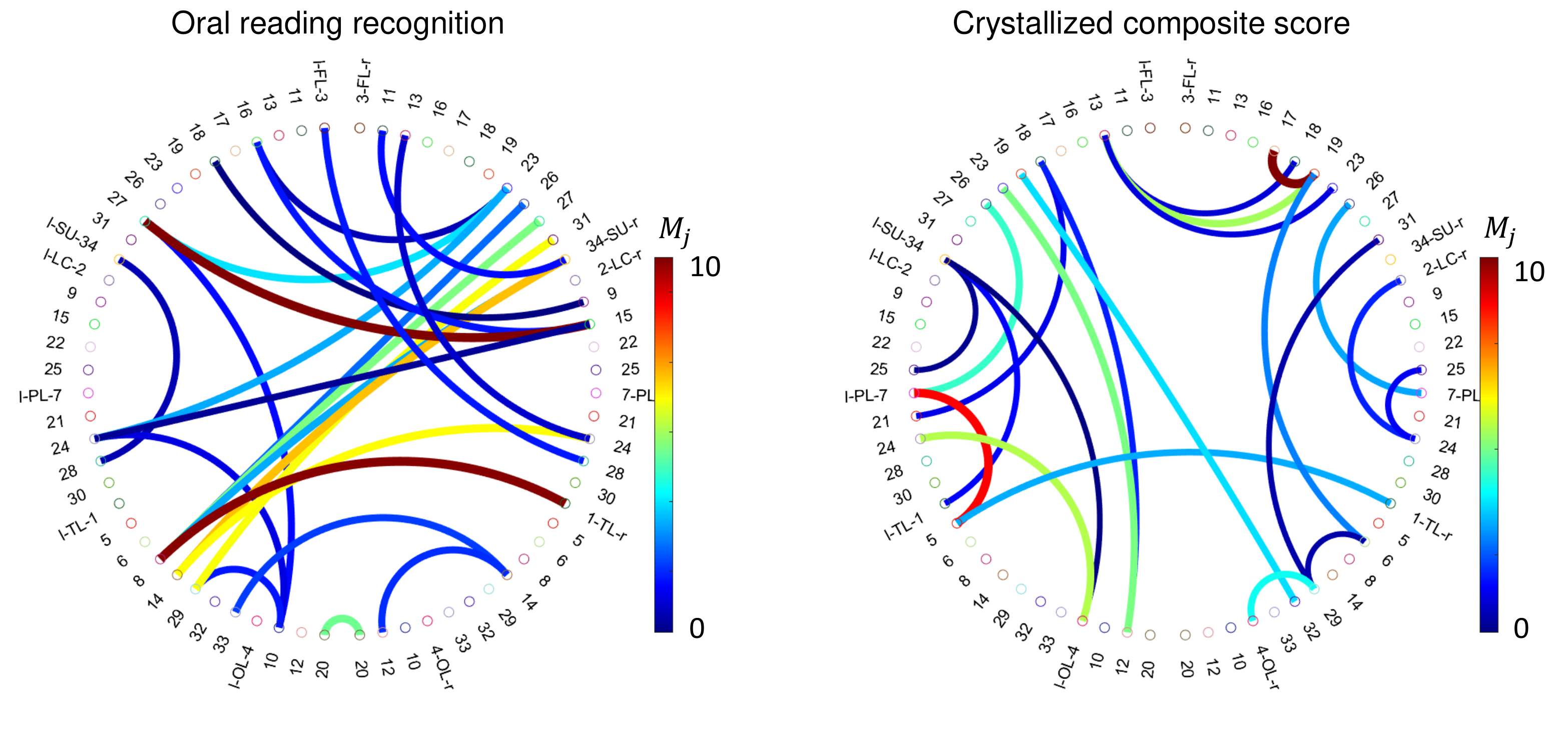}
\caption{Left: selected functional connections associated with Oral reading recognition test. Right: selected functional connections associated with crystallized composition score.}\label{fig:real}
\end{figure*}

As shown in the left panel of Figure \ref{fig:real},
we observe dense connections within both the left and right frontal lobes. In particular, we see that brain regions such as r23 (rostral middle frontal), and l27
(caudal middle frontal) are densely involved (nodes with high degrees) in the selected connections.
We also observe that the four nodes
in the temporal lobe (regions 8, 14, 29), all appear in the selected connections to the frontal lobe. The frontal lobe processes how you interpret language. The temporal lobes are also believed to play an important role in processing auditory information and with the encoding of memory \citep{perrodin2014auditory}. 
Hence, our results show that richer connections between the visual and auditory are strongly associated with higher learning ability in vocabulary. 
As shown in the right panel of Figure \ref{fig:real}, we observe more connections between the front lobe and the temporal lobe.
 Temporal lobe regions are linked to memory, language, and visual perception \citep{allone2017neuroimaging}. Additionally, we observe a high level of communication between the left and right brain hemispheres,
 which contributes to higher cognition
\citep{boisgueheneuc2006functions}.

\section{Conclusion}\label{sec:conclusion}
In this paper, we consider the multiple hypothesis testing procedure which aims at controlling FDR without assuming a model of the
conditional distribution of the response. We combine the idea of data-splitting, Gaussian mirror, and sliced inverse regression (SIR) to construct mirror statistics. With the aid of the developed central limit theorem of SIR, it is shown that the mirror statistic is approximately symmetric with respect to zero when the null hypothesis is true. We then provide an estimator of the false discovery proportion for the single-step testing method and suggest a data-driven threshold based on such an estimator. It is theoretically shown that the FDR of the proposed method is less than or equal to a designated level asymptotically.

The theoretical investigation of the SIR using the newly developed Gaussian approximation theory paves the road for the statistical inference for sufficient dimension reduction with a diverging dimension. It has the potential to be used for other sufficient dimension reduction methods such as sliced average variance estimator (SAVE, \cite{dennis2000save}), principle Hessian directions (PHD, \cite{li1992principal}), and many others. We will leave them in future discussions.

The proposed approach is flexible due to the lean assumptions on the link function. The popular deep neural network could be viewed as the multiple index model. In an ongoing project, it would be interesting to see how the proposed method could be used to reduce the dimension under this framework.

\section{Acknowledgements}
Data used in the preparation of this article were obtained from the Human Connectome Project (www.humanconnectome.org). The HCP WU-Minn Consortium (Principal Investigators: David Van Essen and Kamil Ugurbil; 1U54MH091657) were funded by the 16 NIH Institutes and Centers that support the NIH Blueprint for Neuroscience Research; and by the McDonnell Center for Systems Neuroscience at Washington University. We would like to thank Bing Li, Wen Zhou, Asaf Weinstein for their comments and suggestions. Xing's research is supported by NSF (DMS-2124535).

\bibliographystyle{plainnat}
\bibliography{ref,sir}

\end{document}